\documentclass[final,5p,times,twocolumn,authoryear]{elsarticle}
\usepackage{soul}
\usepackage{graphicx}
\usepackage{dcolumn}
\usepackage{bm}
\usepackage{amsmath}
\usepackage{mathtools}
\usepackage{amssymb}
\usepackage{lipsum}
\usepackage{cuted}
\usepackage{floatflt,epsfig}
\usepackage{xcolor}
\usepackage[normalem]{ulem}

\bibpunct{[}{]}{,}{n}{}{,}  
\usepackage{orcidlink} 
\usepackage{booktabs}

\newcommand{\sat}{\mathrm{sat}}

\newcommand{\cc}[1]{\mathnormal{#1}}

\newcommand{\vnabla}{\boldsymbol{\mathbf\nabla}}

\newcommand{\dash}{---}

\definecolor{mbscolor}{rgb}{0.60, 0.0, 0.65}

\begin{document}

\begin{frontmatter}


\title{Skyrme-Hartree-Fock-Bogoliubov mass models on a 3D mesh: V. The N2LO extension of the Skyrme EDF}


\author[first,second]{G. Grams}
\author[first,third]{W. Ryssens}
\author[first,third]{A. Sánchez-Fernández}
\author[first,third]{N. N. Shchechilin}
\author[first,third]{L. González-Miret Zaragoza}
\author[first,third]{P. Demol}
\author[first,third]{N. Chamel}
\author[first,third]{S. Goriely}
\author[fourth]{M. Bender}

\affiliation[first]{
  organization={Institut d'Astronomie et d'Astrophysique, Université Libre de Bruxelles},
  city={Brussels},
  country={Belgium}
}

\affiliation[second]{
  organization={Institut für Physik und Astronomie, Universität Potsdam},
  city={Potsdam},
  postcode={D-14476}, 
  country={Germany}
}

\affiliation[third]{
  organization={Brussels Laboratory of the Universe -- BLU-ULB},
  city={Brussels},
  country={Belgium}
}

\affiliation[fourth]{
  organization={Université Claude Bernard Lyon 1, CNRS/IN2P3, IP2I Lyon, UMR 5822},
  postcode={F-69100}, 
  city={Villeurbanne},
  country={France}
}

\begin{abstract}
We present BSkG5, the latest entry in the Brussels-Skyrme-on-a-Grid (BSkG) series and the first large-scale nuclear structure model based on next-to-next-to-leading order (N2LO) Skyrme energy density functional (EDF).
By extending the traditional Skyrme EDF ansatz with central terms containing up to four gradients, we are able to combine an excellent global description of nuclear ground state properties with a stiff equation of state for pure neutron matter that is consistent with all astronomical observations of neutron stars. More precisely, the new model matches the accuracy of earlier BSkG models but with two parameters less: we achieve root-mean-square deviations of 0.649 MeV for 2457 atomic masses, 0.0267 fm for 810 charge radii, and 0.43 MeV for 45 primary fission barriers of actinide nuclei. We demonstrate that the complexities of N2LO EDFs
are not insurmountable, even for demanding many-body calculations.
\end{abstract}


\end{frontmatter}

%
\section{Introduction}
\label{sec:intro}

Besides being fascinating objects of study themselves, 
the properties of atomic nuclei are crucial to many areas in physics and beyond: from the search for beyond-standard-model physics to astrophysical simulations, and from reactor design to medical applications. The data needed are both numerous and difficult to obtain: for example, simulations of the rapid neutron-capture process (r-process) require the structural properties, reaction and decay rates of several thousand extremely neutron-rich isotopes. Experimental efforts will not be able to satisfy this need for the foreseeable future, hence nuclear theory needs to fill the data gap with models that can extrapolate to high energy, large angular momentum and/or extreme proton-neutron asymmetry. To make such predictions reliable, models should contain as much of our knowledge of the physics of nuclei as feasible and propagate it to all relevant quantities simultaneously~\cite{Arnould20}.

Today, the most sophisticated tools that can be applied 
to all systems across the chart of nuclei are based on the use of
nuclear energy density functionals (EDFs): these objects link the nuclear binding energy to a variety of mean-field densities that encode the detailed structure of the nucleus in terms of its constituent neutrons and protons. EDF modelling starts by building an ansatz for the analytical form of the EDF in terms of such densities; although formal requirements~\cite{Carlsson08,dobaczewski2000a} put some limits on what terms can appear, the development of such ansätze has so far been mostly guided by empirical considerations. Over the years, three types of analytical forms have come to dominate the literature: Skyrme, Gogny and relativistic EDFs~\cite{Bender03}. Each family has repeatedly demonstrated that they permit parameterizations capable of describing nuclear bulk properties such as radii and masses reasonably  
well\footnote{The D1M Gogny parameterization merits special mention: it is so far unsurpassed in data completeness on nuclear strength functions.~\cite{martini2016,martini2014,goriely2018}.}
\cite{goriely2009a,Kortelainen10,Kortelainen12,pena-arteaga2016,taninah2024,batail2025}, even if easy-to-use Skyrme parameterizations have been pushed the farthest in terms of accuracy on known data and the range of quantities tackled. Nevertheless, several issues continue to plague even state-of-the-art parameterizations. In particular, the predicted single-particle spectra are typically not sufficiently realistic, such that models fail to reproduce several spectroscopic quantities, such as (a) the ground state angular momenta of odd-mass and odd-odd systems~\cite{bonneau2007}, (b) detailed patterns in charge radii such as `kinks'~\cite{reinhard1995,perera2021} and odd-even staggering~\cite{reinhard2017} and (c)  the deformed shell gaps in superheavy nuclei~\cite{Dobaczewski2015}. We also note that the standard EDF forms  (i) have formal issues that prevent their use in beyond-mean-field techniques~\cite{bender2009,duguet2009}, and (ii) find it difficult to reconcile the existence of massive neutron stars (NS) (with a mass $M \geq 2~M_{\odot}$, $M_\odot$ denoting the mass of the Sun) with an accurate global description of nuclear masses~\cite{chamel2009,chamel2011,goriely2009a,batail2025}.

The lack of progress on these issues\dash{}especially in a Skyrme context~\cite{Kortelainen14}\dash{}has motivated the community to develop several new analytical 
forms~\cite{Carlsson08,Raimondi2011,Davesne2014,sadoudi2013,lacroix2015,fayans2000,nakada2013,raimondi2014,bennaceur2017,papakonstantinou2018,grasso2017,burrello2020,bulgac2018,chamel2009,batail2023,zurek2024}. Among these, N$\ell$LO Skyrme EDFs seem particularly promising: these forms feature all possible zero-range terms in which figure up to $2\ell$ derivatives\dash{}the traditional Skyrme form is an NLO EDF in this parlance~\cite{Carlsson08}. On the formal front, N$\ell$LO EDFs (i) naturally arise if one generalizes Skyrme's effective interaction to higher orders of relative momenta~\cite{Raimondi2011,Davesne2014} and (ii) can be used to systematically approximate finite-range 
interactions~\cite{carlsson2010,dobaczewski2010}\dash{}also in computational chemistry~\cite{maximoff2001a}. Nevertheless, it must be said that 
the notion of N$\ell$LO in gradients does not refer to a formal power counting scheme in the sense of an effective field theory.
On a purely phenomenological side, the new terms offer degrees of freedom that\dash{}compared to NLO\dash{}allow for more flexibility to describe infinite nuclear matter (INM)~\cite{Davesne2015} and nucleon optical potentials~\cite{wang2018,wang2024,WangEtAl2025b} and could perhaps be used to 
improve the description of single-particle spectra.

N$\ell$LO forms have not yet been systematically explored in actual simulations of atomic nuclei and extreme astrophysical environments. 
There are several reasons for this: N$\ell$LO EDFs are complex objects, both in practical and numerical 
terms~\cite{becker2015,carlsson2010a,ryssens2019a}, that admit multiple equivalent 
formulations~\cite{Ryssens21} and that contain many more terms that can trigger finite-size 
instabilities than standard NLO EDFs.
The ambition of the adjustment of SN2LO1, the only N2LO parameterization
adjusted to the properties of finite nuclei to date [49, 50], was limited
to the proof of concept, and did not yet systematically improve on predictive
power compared to a standard NLO Skyrme EDF fitted to a similar data set.

In this Letter, we realize part of the potential of N$\ell$LO forms: we present BSkG5, a large-scale model of nuclear structure in the Brussels-Skyrme-on-a-Grid series~\cite{Scamps21}, that is based on an N2LO analytical form. Adjusted on the data of thousands of nuclei, the model offers state-of-the-art global agreement for several nuclear properties and a realistic description of INM that is on par with earlier models in the series. In particular, the N2LO form allows us to reconcile the accuracy for nuclear masses with an equation of state (EoS) of cold dense matter consistent with the existence of massive NSs. Earlier NLO models, such as BSkG3~\cite{Grams23} and BSkG4~\cite{Grams25}, achieved this by expanding the density dependence of the EDFs coupling constants; BSkG5 achieves this without the formal issues associated with such terms~\cite{duguet2009} and with two parameters less. BSkG5 will not be the last word:
one goal of this Letter is the \emph{practical} demonstration that the complexities of N$\ell$LO EDFs can be overcome and that these forms offer a promising path forward.

\section{Formalism \& parameter adjustment}

We describe the nucleus with Bogoliubov states: to such a state we associate a total energy that
accounts for the nucleons' kinetic energy and both the Coulomb and strong interaction between nucleons,
but also approximate corrections for beyond-mean-field correlations. By minimizing the total energy,
we find the single Bogoliubov state from which we consistently compute the properties of a given nucleus.
For details on the modelling of all contributions and our approach to solving the mean-field equations,
we refer the reader to Refs.~\cite{Scamps21,Ryssens22,Grams23,Grams25}.

  The analytical form of the Skyrme EDF is what models the strong interaction between nucleons in the particle-hole channel. We seek now to build a model based on the more general N2LO form, whose contribution to the total energy we
  write as
\begin{equation}
\label{eq:Eskyrme}
E_{\rm Sk} = \int d^3 \bold{r}  \left[
                                            \mathcal{E}_{\text{Sk}}^{(0)}(\bold{r})
                                          + \mathcal{E}_{\text{Sk}}^{(2)} (\bold{r})
                                          + \mathcal{E}_{\text{Sk}}^{(4)} (\bold{r})
                                          \right]  \, ,
\end{equation}
where the energy densities $\mathcal{E}^{(k)}_{\rm Sk}$ collect all EDF terms that involve $k = 0,2,4$ derivatives.
It is useful to decompose these objects further:
$
\mathcal{E}_{\text{Sk}}^{(i)} (\bold{r}) =  \sum_{t = 0,1} \left[
                                            \mathcal{E}^{(i)}_{t, \rm e}(\bold{r})
                                          + \mathcal{E}^{(i)}_{t, \rm o}(\bold{r})
                                          \right]  \,
$
where $t=0,1$ is an isospin index and the subscript e(o) indicates whether the local densities from which the EDF terms are constructed are even (odd) under time reversal. All terms in traditional Skyrme models would be part of the first two terms in Eq.~\eqref{eq:Eskyrme}; we choose the LO and NLO parts of our EDF in a rather standard way
and (a) do not incorporate the rarely-used explicit NLO tensor force \cite{lesinski2007}, but
(b) adopt the widely-used extended spin-orbit prescription of Refs.~\cite{reinhard1995,sharma1995}, 
and (c) consistently keep all spin-gradient terms from the central interaction, despite their tendency to induce finite-size instabilities.
The new part of the EDF is $\mathcal{E}^{(4)}_{\rm Sk}$; following the conventions of Ref.~\cite{Ryssens21}, the N2LO time-even energy density we adopt reads
%
\begin{align}
\label{eq:SkTeven:4}
  \mathcal{E}^{(4)}_{\text{t,e}} (\bold{r})
  = & {
     \cc{A}^{(4,1)}_{t,\textrm{e}} \big(\Delta D^{1,1}_t \big) \big( \Delta D^{1,1}_t \big)
     }
  + \cc{A}^{(4,2)}_{t,\textrm{e}} D^{1,1}_t \, D^{\Delta, \Delta}_t \\
   +& \cc{A}^{(4,3)}_{t,\textrm{e}} D^{(\nabla, \nabla)}_t D^{(\nabla, \nabla)}_t
   + \cc{A}^{(4,4)}_{t,\textrm{e}} \sum_{\mu\nu} D^{\nabla, \nabla}_{t, \mu\nu}
                                                 D^{\nabla, \nabla}_{t, \mu\nu}
\nonumber \\
   +& \cc{A}^{(4,5)}_{t,\textrm{e}} \sum_{\mu\nu} D^{\nabla, \nabla}_{t,\mu\nu}
                                              \big( \nabla_{\mu} \nabla_{\nu}  D^{1,1}_t \big)
    {
   +  \cc{A}^{(4,6)}_{t,\textrm{e}} \sum_{\mu\nu} C^{1,\nabla\sigma}_{t,\mu \nu} 
      \big( \Delta C^{1,\nabla\sigma}_{t,\mu \nu} \big)
}
 \nonumber \\
   +& {
     \cc{A}^{(4,7)}_{t,\textrm{e}}  \sum_{\mu\nu\kappa} 
            \big(       \nabla_{\mu} C^{1,\nabla\sigma}_{t,\mu \kappa} \big)
            \big(       \nabla_{\nu} C^{1,\nabla\sigma}_{t,\nu \kappa} \big)
}
    {
   +  \cc{A}^{(4,8)}_{t,\textrm{e}} \sum_{\mu\nu} 
       C^{1,\nabla\sigma}_{t,\mu\nu} C^{\Delta,\nabla \sigma}_{t,\mu\nu}
       } \, \nonumber . 
\end{align}

The local densities that form this expression are denoted as $C^{A,B}_t$ and $D^{A,B}_t$, with an isospin index $t$ and superscripts that indicate their operator structure. The numerous coupling constants $A^{(i,j)}_t$ can be completely specified in terms of just four parameters $t_1^{(4)}, t_2^{(4)}, x_1^{(4)}$ and  $x_2^{(4)}$~\cite{Ryssens21}. This form of $\mathcal{E}^{(4)}_{\rm Sk}$ corresponds to the \emph{central} N2LO extension of a zero-range interaction~\cite{Raimondi2011} and does not feature N2LO spin-orbit or tensor terms.
We provide the complete EDF form, including the $\mathcal{E}^{(4)}_{\rm t, o}$ terms, and details on local densities and coupling constants, as well as the adopted microscopic pairing in the supplementary material~\ref{app:supl}. 

We constructed BSkG5 in close analogy to BSkG4, adopting\dash{}aside from the EDF form\dash{}the same modelling ingredients and fitting protocol: while imposing realistic INM properties and ensuring a maximal NS mass above $2~M_{\odot}$, we adjusted the model to essentially all known nuclear masses, charge radii and the fission properties of 45 actinide nuclei using a three-dimensional and numerically accurate representation of the nucleus that allows for both triaxial and octupole deformation as well as time-reversal symmetry breaking~\cite{Scamps21}. Incorporating the N2LO extension into this large-scale parameter adjustment proved challenging. First, we soon realized that a sufficiently stiff EoS is only feasible when the coupling constants $\cc{A}^{(4,2)}_t$ 
that multiply the term containing the higher-order kinetic density $D^{\Delta, \Delta}_t$
take positive values since the corresponding EDF terms dominate the energy of INM at high density. Second, the issue of finite-size instabilities\dash{}likely worsened compared to NLO EDFs due to the higher-order 
derivatives\dash{}excluded large portions of the parameter space.
Although the linear response of INM can offer some clues~\cite{hellemans2013,pastore2015}, no procedure can 
ensure the absence of such instabilities with certainty. We adopted the pragmatic approach of relying on the sensitivity of our coordinate-space calculations to such problems to avoid unstable regions of parameter space; we document the stability of BSkG5 below and in the supplementary material. Nevertheless, we cannot guarantee the stability of the model up to infinite transferred momenta; for future uses of BSkG5, we advocate an ultraviolet cutoff at a mesh spacing of $0.4$ fm or $k \sim 7.9 $ fm$^{-1}$;
the latter significantly exceeds momenta relevant to the description of ground state nuclear properties and NS.

\section{Results}
\label{sec:res}

\subsection{Properties of atomic nuclei}

\begin{figure}[h]
\begin{center}
\includegraphics[width=0.9\columnwidth]{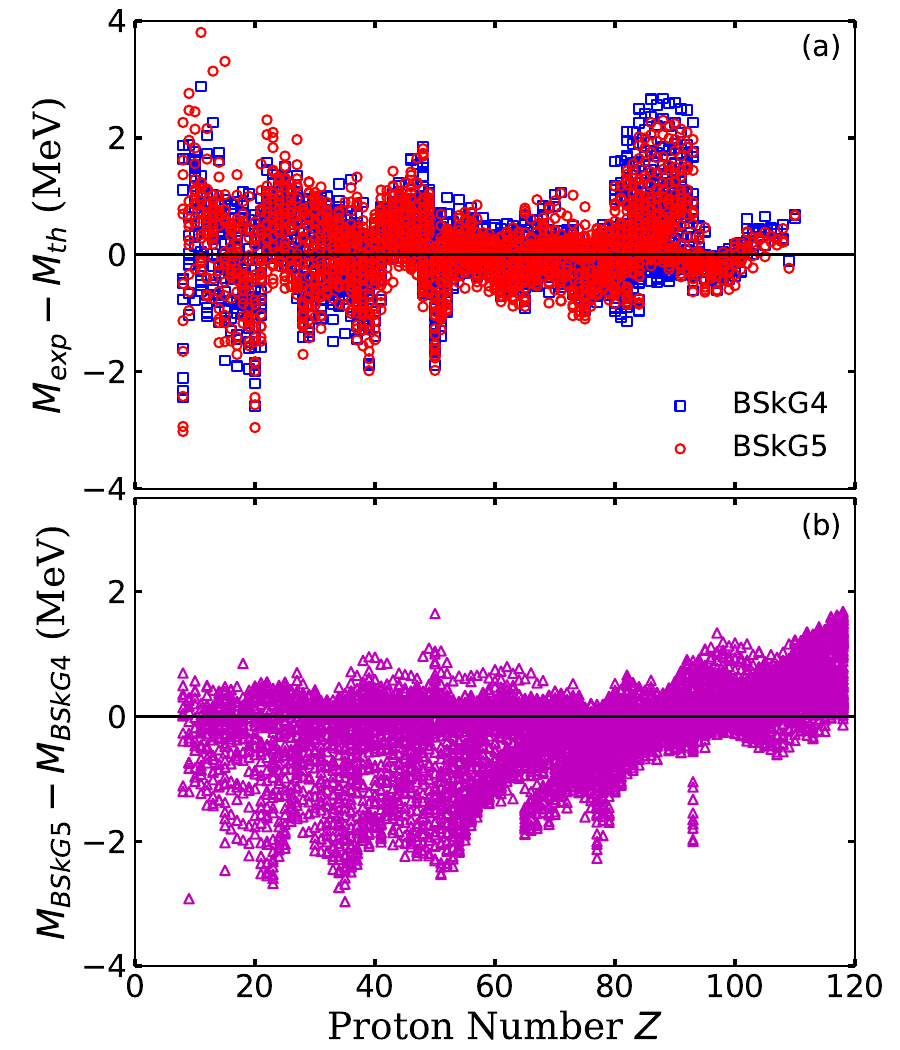}
\end{center}
\caption{\label{fig:massexp} 
Panel (a): Differences between experimental \cite{AME2020} and calculated masses for BSkG5 (red circles) and BSkG4 (blue squares) masses, 
where positive values indicate overbound nuclei.
Panel (b): Mass differences between BSkG5 and BSkG4 for all nuclei with $8\le Z \le 118$ lying between the BSkG5 proton and neutron drip lines.
} 
\end{figure}

Table~\ref{tab:rms} showcases the new models' global performance for nuclear properties\dash{}as compared to BSkG4\dash{}through the mean $(\bar{\epsilon})$ and root-mean-square (rms) ($\sigma$) deviations for different quantities. BSkG5 arrives at a global description of the known masses ($\sigma = 0.643$ MeV) and their differences ($S_n, S_p, Q_{\beta}, \Delta^{5}$) in AME20~\cite{AME2020}---see Ref.~\cite{Scamps21} for their definitions---that is quantitatively similar to its predecessor. The top panel of Fig.~\ref{fig:massexp} shows the difference between calculated and experimental masses for BSkG4 and BSkG5: the overall behaviour of the error as a function of $Z$ and $N$ (not shown) is similar. The bottom panel of Fig.~\ref{fig:massexp} shows the difference between the mass tables of both models, with differences generally limited to 2 MeV and 3 MeV for outliers; while such a difference remains sizeable from the point of view of applications, they are in fact remarkably small for two models based on a different EDF form and in line with differences between older BSkGs~\cite{Grams25}. The performance of BSkG5 w.r.t. masses is not significantly better than its predecessor, or many of the BSk models for that matter~\cite{Goriely16}, but its mass rms is much lower than that of many other available Skyrme NLO parameterizations~\cite{Kortelainen14}. In particular, it is much lower than the rms of SN2LO1, which we evaluated at $\sigma(M) = 6.08$ MeV for 866 even-even nuclei.
\footnote{SN2LO1 does not specify pairing terms~\cite{Becker17}\dash{}we employed the pairing terms used in Ref.~\cite{Ryssens21} to obtain this estimate for $\sigma(M)$ for SN2LO1 but did not employ the Lipkin-Nogami prescription.}

\begin{table}[]
    \centering
\caption{\label{tab:rms} rms 
         ($\sigma$) and 
         mean  ($\bar \epsilon$) deviations between experimental and calculated values for the BSkG4 and BSkG5 models. 
         The first block refers to the nuclear ground-state properties, and the second and third ones to fission properties. More specifically, these 
         values were calculated with respect to 2457 known masses ($M$)~\cite{AME2020} 
         of nuclei with $Z$, $N \geq 8$, the subset of 299 known masses $M_{nr}$ of neutron-rich nuclei with $S_n \le 5$~MeV, the subset of 22 known masses of nuclei that are one nucleon off from being doubly magic ($M^{\rm magic}_{A\pm1}$), the 1784 (2109) five-point neutron (proton) gaps $\Delta^{5}_{n}$ $(\Delta^{5}_{p})$, 2309 neutron separation energies 
         ($S_n$), 2173 $\beta$-decay energies ($Q_\beta$),
         810 measured charge radii
         ($R_c$) \cite{Angeli13}, 45 empirical values for primary ($E_{\rm I}$)
         and secondary ($E_{\rm II}$) fission barrier heights~\cite{Capote09}, 28 fission isomer excitation energies ($E_{\rm iso}$) of actinide nuclei~\cite{Samyn04} and 107 ground state SF half-lives \cite{Kondev2021}. The first line reports the model error \cite{Moller88} for all measured masses. }
    \begin{tabular}{c|cc}
        \hline
        \hline
Results & \mbox{BSkG4} & \mbox{BSkG5} \\
\hline
$\sigma_{\rm mod}(M)$ [MeV]           &  0.629   &   0.643 \\ 
$\sigma(M)$ [MeV]                     &  0.633   &  0.649 \\ 
$\bar \epsilon (M)$ [MeV]             &  +0.104  &  +0.085  \\ 
$\sigma(M_{\rm nr})$ [MeV]            &  0.672   &  0.751   \\ 
$\bar \epsilon (M_{\rm nr})$ [MeV]    &  +0.037  &  +0.182  \\ 
$\sigma(M^{\rm magic}_{\rm A\pm1})$ [MeV]                    & 1.082  &  0.952\\
$\bar \epsilon(M^{\rm magic}_{\rm A\pm1})$ [MeV]                    & -0.007  &  -0.157 \\
$\sigma(\Delta_n^5)$ [MeV]            &   0.290     &  0.291  \\ 
$\bar \epsilon (\Delta_n^5)$ [MeV]    &   -0.041    & +0.039   \\ 
$\sigma(\Delta_p^5)$ [MeV]            &   0.427     &  0.421  \\ 
$\bar \epsilon (\Delta_p^5)$ [MeV]    &  +0.005     & +0.008   \\
$\sigma(S_n)$ [MeV]                   &  0.402   &  0.409 \\ 
$\bar \epsilon (S_n)$ [MeV]           & +0.010   &  +0.003  \\ 
$\sigma(Q_\beta)$ [MeV]               & 0.493    &  0.526  \\ 
$\bar \epsilon (Q_\beta)$ [MeV]       & +0.026   &  +0.007   \\ 
$\sigma(R_c)$ [fm]                    & 0.0246   &   0.0267 \\ 
$\bar \epsilon (R_c)$ [fm]            & +0.0006  &  +0.0007   \\ 
\hline
$\sigma(E_{\rm I})    $  [MeV]         &   0.36  &   0.43    \\
$\bar{\epsilon}(E_{\rm I})  $  [MeV]   &  -0.02  &  -0.02     \\
$\sigma(E_{\rm II})   $  [MeV]         &   0.53  &  0.49     \\
$\bar{\epsilon}(E_{\rm II}) $  [MeV]   &  -0.02  &  +0.10     \\
$\sigma(E_{\rm iso})  $  [MeV]         &   0.33  &  0.59     \\
$\bar{\epsilon}(E_{\rm iso})$  [MeV]   &  +0.05  &  -0.12 \\
\hline
$\sigma'(t_{1/2}^{\rm SF})    $           &   $3.7\cdot 10^{4}$  &   $1.3\cdot 10^{3}$    \\
$\bar{\epsilon}'(t_{1/2}^{\rm SF})  $     &  $1.4\cdot 10^{2}$  &  $1.6\cdot 10^{0}$     \\
\hline
\hline
\end{tabular}
\end{table}

BSkG5's performance for masses being similar to those of BSkG1--4 does not mean that the N2LO terms constitute but a small correction. In practice, including N2LO terms forces a significant renormalization of the coupling constants of the LO and NLO terms. Nevertheless, a hierarchy of terms does seem to emerge: for $^{208}$Pb, the N2LO contribution to the binding energy is about $109$ MeV while the LO and NLO contributions amount to  $-6818$ and $355$ MeV, respectively.

The analysis of quantities other than masses also leads to the conclusion that BSkG5 achieves a global description of roughly the same quality as BSkG4. The new model describes accurately the available data on charge radii ($\sigma(R_c) = 0.0267$ fm)\footnote{We consider only 810 experimental charge radii of Ref.~\cite{Angeli13} and exclude the Re, Po, Rn, Fr, Ra, Am, and Cm values derived from systematics.}, empirical fission barriers of actinide nuclei from RIPL-3 ($\sigma(E_{I}) = 0.43$ MeV)~\cite{Capote09}, known spontaneous fission lifetimes (logarithmic $\sigma'(t^{\rm SF})_{1/2} = 1.3 \cdot 10^3$) as well as the rotational moments of inertia of even-even nuclei (see the supplemental material). We refer the interested reader to Refs.~\cite{Scamps21,ryssens2023,Sanchez2025} for more details on how we model these nuclear properties. Comparing in more detail to BSkG4, we note that BSkG5 performs slightly worse for static fission properties\dash{}the isomer excitation energies in particular\dash{}but does better for the spontaneous fission lifetimes, which are also sensitive to \emph{dynamical} aspects of the process~\cite{baran1981}; nevertheless, it does not surpass BSkG3 in accuracy~\cite{Sanchez2025}, see the supplemental material.

\begin{figure}[]
\begin{center}
\includegraphics[width=1.0\columnwidth]{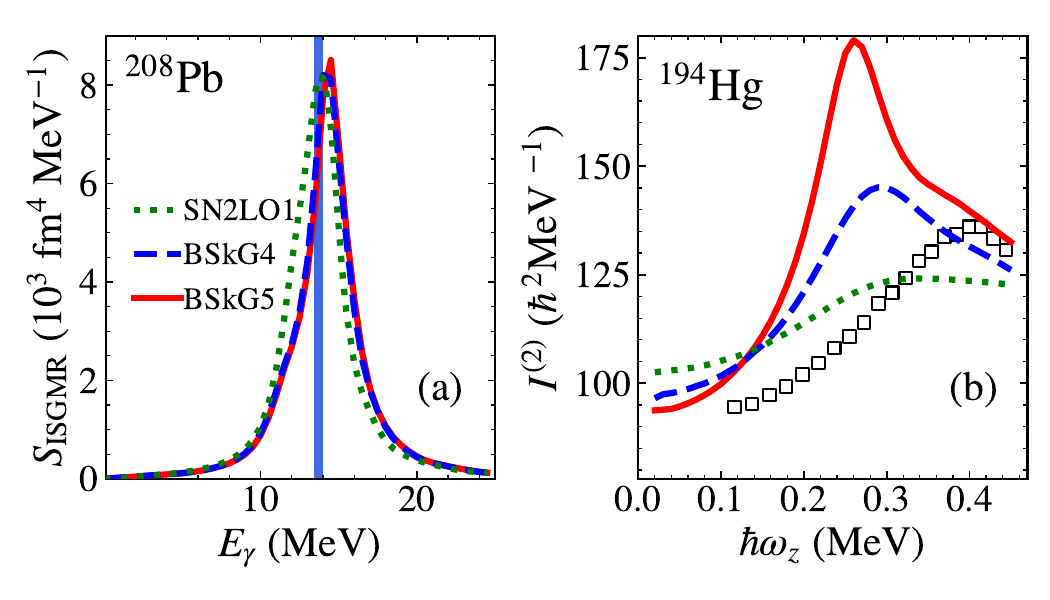}
\end{center}
\caption{\label{fig:LR} 
    Panel (a): FAM photoabsorption strength function of the isoscalar giant monopole resonance  of ${}^{208}$Pb calculated with SN2LO1 (dotted green), BSkG4 (dashed blue), and BSkG5 (solid red). The experimental energy-centroid of the resonance is represented by the blue vertical bar~\cite{patel2013testing}. Panel (b): Dynamical moment of inertia of the superderformed band of ${}^{194}$Hg calculated with the same models. Experimental data (empty squares) are those of the SD-1 band in Ref.~\cite{singh2002}.}
\end{figure}

Much like BSkG4 and previous BSk(G) models, the error of the masses calculated with BSkG5 remains highly correlated with magic numbers and shell structure. The new model offers a rather similar description of the shell structure to many NLO models. For example, the rms on the masses of 22 nuclei that are one nucleon off from being doubly-magic is  $\sigma(M^{\rm magic}_{A\pm1}) = 0.952$ MeV. This is comparable to the value of BSkG4 ($1.082$ MeV) and to earlier BSkG models, and lower than that reported for other Skyrme parameterizations~\cite{Kortelainen14}. This deviation is significantly larger than the model's global rms and indicates that the model imperfectly captures the structure of nuclei close to magicity. In the supplemental material, we show that the single-particle spectrum of $^{208}$Pb predicted by BSkG5 is close to those of BSkG1--4. Despite the hope placed in N2LO EDFs regarding shell structure, the absence of a reshuffling of the single-particle spectrum should not be taken as disappointing for two reasons: (a) nothing in our optimisation protocol targeted single-particle structure directly and (b) non-central terms that could be finetuned for this purpose are not included in Eq.~\eqref{eq:SkTeven:4}.

  Despite the retention of spin-gradient terms and the terms with higher order gradients, BSkG5 seems
    to be free of finite-size instabilities for all practical intents and purposes. Although we encountered
    many such instabilities throughout the parameter adjustment, the final parameterization led to converged
    calculations for thousands of ground states. In particular, we emphasize that (a) a coordinate-space representation such as ours is well-suited to detect such instabilities~\cite{hellemans2013} and (b) our calculations for odd-mass and odd-odd nuclei naturally included the time-odd terms of the EDF, meaning that
    the majority of our calculations could have detected spin instabilities that would have remained hidden for time-reversal conserving calculations. We discuss these matters in more detail in the supplementary material, but demonstrate here with two more illustrative examples--in addition to the fission calculations already mentioned--that BSkG5 gives stable results even when going beyond ground-state properties. Fig.~\ref{fig:LR} shows for SN2LO1, BSkG4 and BSkG5 (a) the isoscalar monopole strength function of $^{208}$Pb and (b) the dynamical moment of inertia $\mathcal{I}^{2}$ along the superdeformed (SD) rotational band in $^{194}$Hg as a function of rotational frequency. We obtained the monopole strength function from finite amplitude method (FAM) for random-phase approximation (RPA) 
    calculations in our 3D numerical representation along the lines of Ref.~\cite{washiyama2017}: our simulations included all particle-hole excitations up to 60 MeV while consistently accounting for all ingredients of the corresponding models; the resulting strength functions exhaust the energy-weighted sum rule for more than 99.4\% in all three cases. BSkG4 and BSkG5 produce centroid energies of the giant monopole resonance--$E_{\rm cent} = 14.3$ and $14.4$ MeV--that are somewhat larger than the experimental value $E_{\rm cent} = 13.7 \pm 0.2$ MeV~\cite{patel2013testing}. 
This result is not unexpected. It is to be recalled first that $K_{\nu}$ is to a large extent
fixed by a correlation with other nuclear matter properties, first identified for standard 
NLO EDFs in Ref.~\cite{chabanat1997} and generalised for N2LO EDFs in Ref.~\cite{Becker17}
Without a dedicated analysis\dash{} outside the scope of this
letter\dash{}it is however not clear if the correlation between $E_{\rm cent}$ of the GMR 
and $K_{\nu}$ that establishes the empirical value~\cite{colo2004} remains the same when
using a NLO EDF with extended density dependence or a N2LO EDF instead of a standard NLO EDF.
In view of this, it is gratifying to see that BSkG4 and BSkG5 describe $E_{\rm cent}$ of
$^{208}$Pb reasonably well without having been finetuned to that observable.
The dynamical moment of inertia of the SD band in $^{194}$Hg 
is obtained through cranked HFB calculations as in Refs.~\cite{Ryssens21,Ryssens22}\footnote{Our results for SN2LO1 differ from those in Ref.~\cite{Ryssens21} because we omitted here the Lipkin-Nogami prescription in favor of the stabilisation recipe of Ref.~\cite{erler2008} - but only for our calculations of the SD band.}. The deviation of BSkG5 w.r.t.\ the experimental $\mathcal{I}^{(2)}$ is largest among the models we consider here, but it is nevertheless reasonable given the sensitivity of this type of 
observable
to details of the EDF~\cite{hellemans2012}. Although the description of the data is not remarkable for either the monopole strength or the SD rotational band, these calculations go significantly beyond our ground state calculations\dash{}for instance, both are sensitive to time-odd terms with the cranked calculations even reaching angular momenta above $50 \hbar$\dash{}and demonstrate the feasibility of using N2LO parameterizations even in advanced applications.

\subsection{Nuclear matter and neutron star properties}
\label{sec:NMNS}

\begin{figure}[htbp]
\begin{center}
\includegraphics[width=1.0\columnwidth]{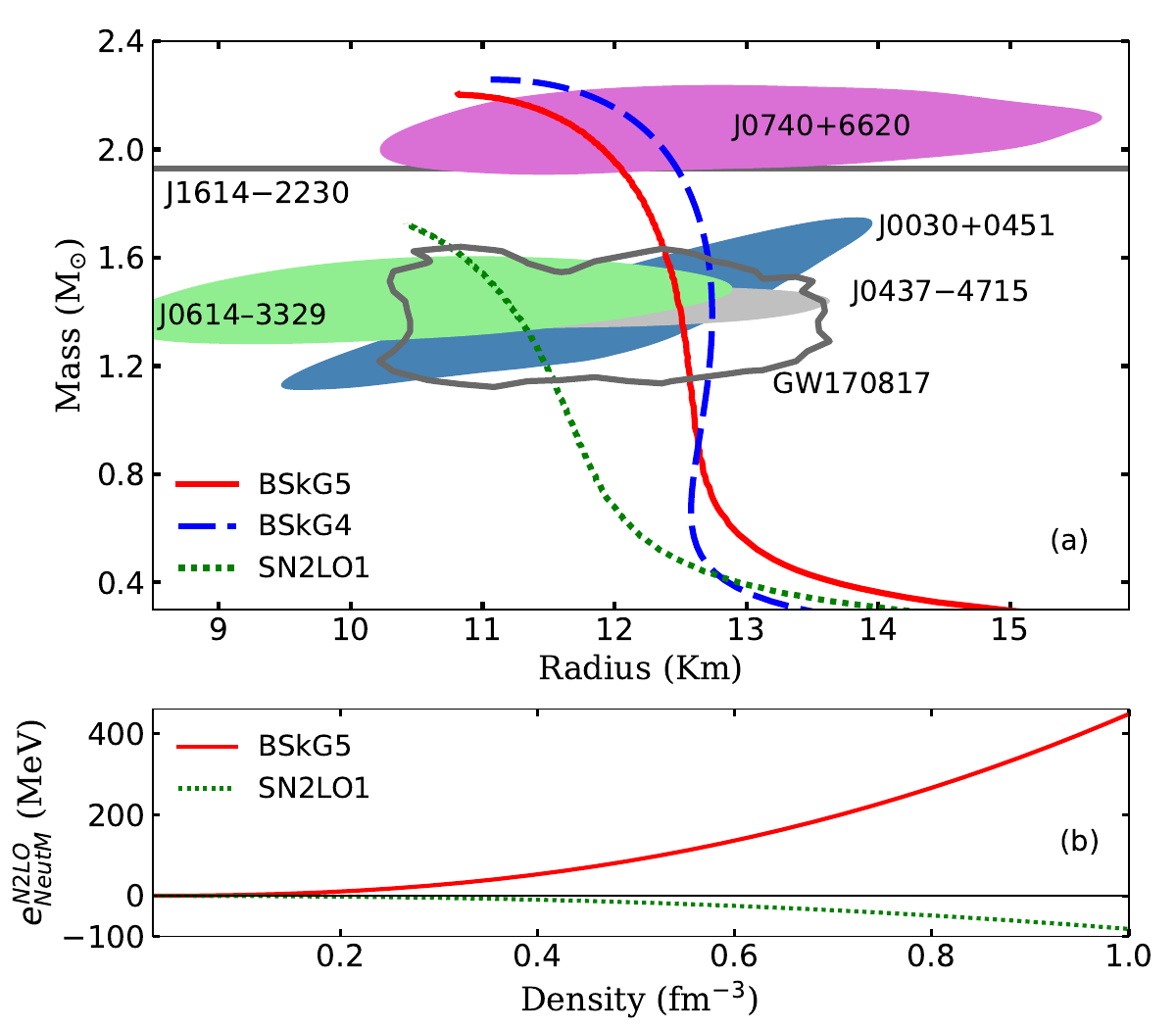}
\end{center}
\caption{
\label{fig:NS} 
Panel (a): Mass-radius curves for families of NS constructed with BSkG5 (solid red), BSkG4 (dashed blue), SN2LO1 (dotted green). Filled contours show the pulsar observations from the NICER telescope, 
PSR J0614–3329~\cite{Mauviard2025}, 
PSR J0030+0451~\cite{Vinciguerra2024},
PSR J0437–4715~\cite{Choudhury2024},
and PSR J0740+6620~\cite{Salmi2024}.
The open contour shows the gravitational wave GW170817 observation from LIGO-Virgo interferometers~\cite{LIGOScientific:2018cki}. The gray band marks the mass measurements of PSR~J1614$-$2230~\cite{Agazie23}.
Panel (b): Contribution from N2LO terms to the energy per particle of pure neutron matter with respect to baryon density for BSkG5 (solid red), and SN2LO1 (dotted green).}
\end{figure}

We show in Table~\ref{tab:inm} the INM properties at saturation for SN2LO1, BSkG4, and BSkG5; we provide a plot of the EoS as well as the corresponding expressions valid for N2LO EDFs in the supplementary material. The INM properties of the new model are within current uncertainties~\cite{Margueron2018a} and comparable to those of BSkG4; the differences in the Landau parameters $G_0, G_0'$ reflect our retention of NLO spin-gradient terms that were omitted for BSkG4. 
BSkG5 improves on SN2LO1 in two ways. First, we obtain the hierarchy of nucleon effective masses $ m^*_n/m > m_p^*/m$ in 
pure neutron matter (NeutM) up to about twice saturation density 
as predicted by more realistic INM calculations (see, e.g., Ref.~\cite{Zuo2006}) in contrast to SN2LO1. 
Second, BSkG5 produces a much stiffer EoS compared to SN2LO1, as we imposed during the parameter adjustment. We illustrate the latter with Fig.~\ref{fig:NS}: panel (a) shows the NS mass-radius curves based on the EoS for cold neutron-proton-electron-muon matter in $\beta$-equilibrium as obtained from Tolman-Oppenheimer-Volkoff (TOV) equations~\cite{Tolman1939,Oppenheimer1939}~\footnote{We used the approximations of Ref.~\cite{Zdunik17} to account for the NS crust. For more details on the EoS and TOV equations, see Refs.~\cite{Pearson18,Grams23,Zdunik17}.}. As BSkG4 before it, BSkG5 is consistent with all constraints arising from (i) the observation of massive pulsars~\cite{Demorest2010,Saffer2025} as the predicted maximum NS mass is 2.2$M_{\odot}$, (ii) the NICER observations~\cite{Mauviard2025, Vinciguerra2024, Choudhury2024, Salmi2024} and (iii)
the observation of GW170817~\cite{LIGOScientific:2018cki}. This agreement does not come at the expense of causality: BSkG5 predicts a sound speed that is smaller than the speed of light throughout the entire NS interior.
The reason why BSkG5 combines success for NS and nuclear properties is its N2LO form; we show in panel (b) the
contribution of the N2LO terms to the energy per particle in pure neutron matter ($e^{\rm N2LO}_{\rm NeutM}$). At large baryon density, the N2LO terms, mostly those of the form $D^{1,1}_t D^{\Delta, \Delta}_t$, 
provide a large positive contribution to the pressure of INM; because the coupling constants $A^{(4,2)}_t$ of SN2LO1 have opposite sign, the latter model does not produce a mass-radius curve consistent with observations. 
To end this section, we note that BSkG5 predicts that the direct Urca process occurs at a baryon density of 0.456~fm$^{-3}$, corresponding to a NS mass of about 1.44~$M_\odot$, as is necessary to interpret the observed thermal luminosity of some NSs~\cite{Lattimer91,Burgio21,Marino_ea24}.

\begin{table}[]
    \centering
\caption{\label{tab:inm} INM properties for the SN2LO1~\cite{Becker17}, BSkG4~\cite{Grams25} and BSkG5 models.
From top to bottom: the Fermi momentum, saturation density, energy per particle of symmetric nuclear matter (SNM), symmetry energy, slope of symmetry energy, neutron effective mass in SNM and pure neutron matter (NeutM), incompressibility parameter, isovector incompressibility, isoscalar skewness, and the last two lines show Landau parameters of SNM.
  See Refs.~\cite{Goriely16,Margueron02,Chamel10} for the various definitions.}
    \begin{tabular}{c|ccc}
        \hline
        \hline
Properties           & \mbox{SN2LO1} & \mbox{BSkG4} & \mbox{BSkG5} \\
\hline
$k_F$~[fm$^{-1}$]          &  1.339   & 1.327   & 1.330  \\
$\rho_{\sat}$~[fm$^{-3}$]  &  0.1620  & 0.1578   & 0.1589 \\
$a_v$~[MeV]                & -15.95   & -16.08   & -16.15 \\
$J$~[MeV]                  &  31.96   & 31.00    & 32.04 \\
$L$~[MeV]                  &  48.89   & 55.67    & 56.82 \\
$(m^*/m)_n^{\rm SNM}$      &  0.709   & 0.860    & 0.860 \\
$(m^*/m)_n^{\rm NeutM}$    &  0.698   & 1.053    & 0.883 \\
$K_v$~[MeV]                &  221.87  & 242.57   & 244.15 \\
$K_{\text{sym}}$~[MeV]     & -126.77   & -19.48   & -90.76 \\
$K^\prime$~[MeV]           &  408.7   &  303.9   & 343.5 \\
$G_0$                      &  1.0501  & 0.2179   & -0.5837  \\
$G_0^\prime$               & -0.0658 & 0.9789   &  0.4747 \\
\hline
\end{tabular}
\label{tab:nucmatter} 
\end{table}

\subsection{Application to the r-process nucleosynthesis in NS mergers}
\label{sec:rpro}

\begin{figure}[htbp]
\begin{center}
\includegraphics[width=1.0\columnwidth]{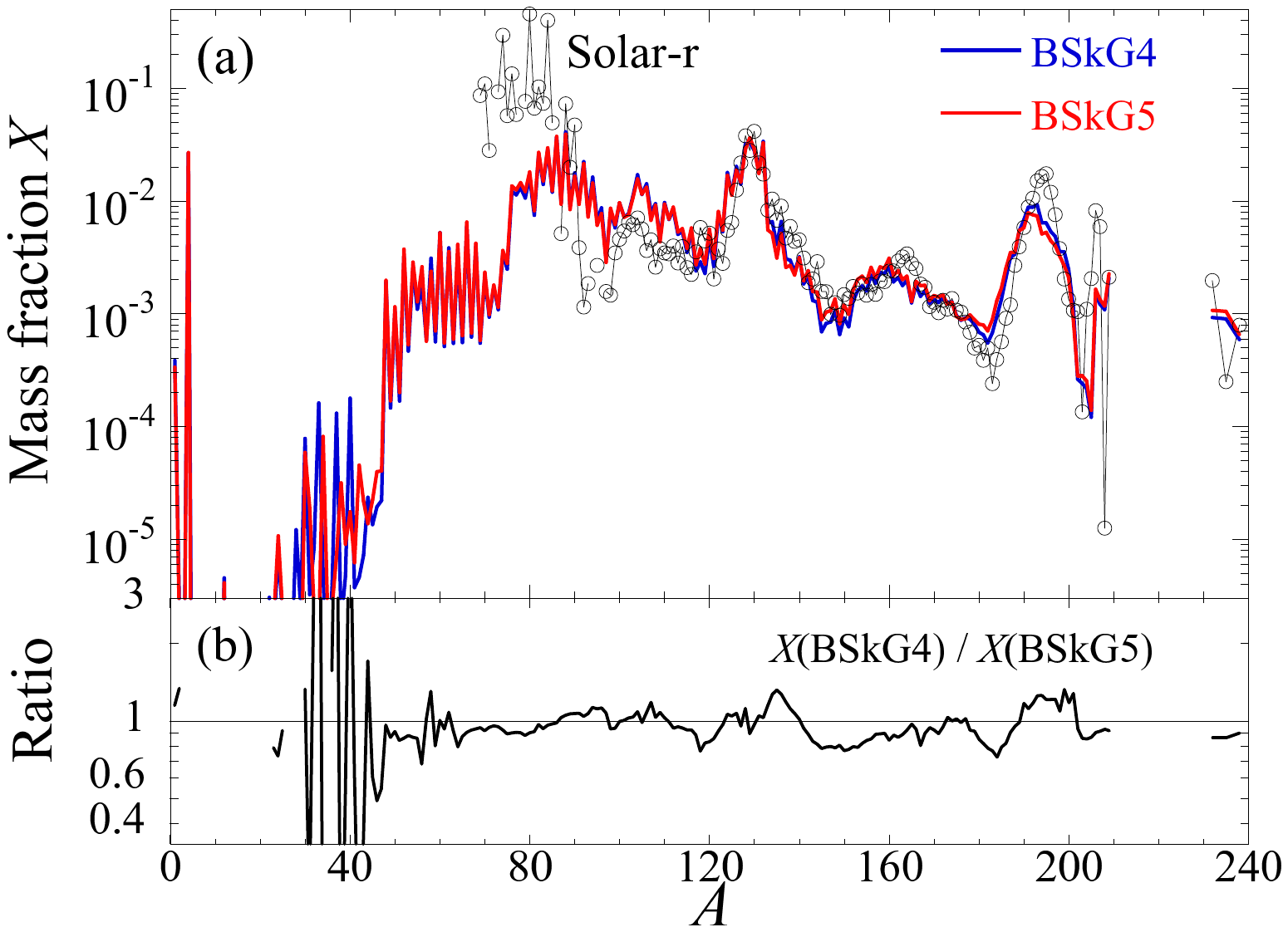}
\end{center}
\caption{Panel (a) Final mass fractions $X$ of the material ejected from a 1.2--1.6 $M_\odot$ NS–NS binary system of Ref.~\cite{Just23}, obtained with the
BSkG4 and BSkG5 masses. The solar system r-abundance
distribution (open circles) from Ref.~\cite{Goriely99} is shown for comparison and 
suitably
normalised. Panel (b) shows the ratio between the predicted BSkG5 and BSkG4 mass fractions for $X>10^{-6}$.}
\label{fig:rpro} 
\end{figure}

We illustrate the impact of the new model on r-process nucleosynthesis in NS mergers in Fig.~\ref{fig:rpro}, where the composition of
the matter ejected from the specific end-to-end simulation of a 1.2--1.6 $M_\odot$ NS-NS binary system~\cite{Just23} (the so-called asym-n1a6 model) is shown. The final abundance distributions are calculated
using the radiative neutron capture and photoneutron emission rates obtained consistently from the BSkG4 and BSkG5 neutron separation energies, but the same $\beta$-decay rates from Ref.~\cite{Marketin16} and fission properties of Refs.~\cite{Grams23,Sanchez2025}. Globally, the mass models give rise to abundance distributions that agree within typically 30\%, except for the underproduced $A\simeq 40$ nuclei.
Both predicted abundances match relatively well the solar system r-distribution for nuclei with $A\ge 80$. This comparison shows that the BSkG mass predictions are relatively consistent from one model to the next and do not lead to major differences w.r.t. astrophysical observables, such as the composition of ejected material from NS mergers or the radioactive decay heat from such events (not shown).

\section{Conclusions \& perspectives }

We have presented BSkG5, the first large-scale nuclear structure model based on an N2LO Skyrme EDF, including terms that include up to four derivatives. With this form of the EDF, we can combine
 an excellent global description of a large range of nuclear properties \dash{} masses, charge radii, fission barriers, and half-lives \dash{} with a neutron-matter EoS that is sufficiently stiff to accommodate the existence of massive pulsars. This feat is impossible with a standard Skyrme EDF, and earlier models such as BSkG3 and BSkG4 managed this by introducing additional density-dependent terms. Compared to these, BSkG5 requires two less parameters but remains competitive in all respects.
 
BSkG5 also demonstrates that complex N2LO EDFs can replace standard Skyrme EDFs in many, if not all, applications from a practical point of view; based on an N2LO EDF we have (i) performed large numbers of symmetry-broken ground state calculations sensitive to the time-odd channel of the EDF, (ii) obtained over a hundred spontaneous fission lifetimes, (iii) solved the linear response of $^{208}$Pb w.r.t.\ isoscalar monopole excitations and (iv) followed a superdeformed rotational band up to high spin. These and other calculations indicate that BSkG5 is stable w.r.t.\ finite-size instabilities for all practical intents and purposes, although we recommend a momentum cutoff in the spirit of effective field theory above $k \sim 7.9 $ fm$^{-1}$.

This work opens several avenues of further research. On the side of nuclear structure, BSkG5 can serve as an anchoring point for parameter adjustments to further explore the N2LO parameter space. For now, we do not advocate moving to N$3$LO or beyond but rather propose the introduction of the N2LO spin-orbit terms. The latter do not contribute to INM and are simultaneously sensitive to the angular momenta of single-particle orbitals~\cite{proust2023}; the inclusion of such terms in the EDF \dash{} and suitable spectroscopic observables in the objective function \dash{} could offer a way to improve the spectroscopic quality of large-scale models while not spoiling the description of nuclear bulk properties and the NS EoS that we have achieved here.

In contrast to NLO Skyrme EDFs, the N2LO form of BSkG5 also implies a momentum- and temperature-dependent nucleon effective mass, 
a characteristic naturally found for Gogny-type EDFs~\cite{Sellahewa12,Ventura92} and in ab-initio studies~\cite{Hassaneen04,Shang20}.
This feature thus might open the way to a more realistic description of thermal effects in INM and applications in hot and dense environments, where the effective mass entering the kinetic energy term largely governs thermal responses~\cite{Fantina2012,Raduta_ea21,Yasin_ea20,Schneider_ea19,Constantinou_ea14,Constantinou_ea15,Raithel_ea21,Fields_ea23}. The behavior of BSkG5 INM at finite temperatures, and its implications in relevant astrophysical environments such as core-collapse supernovae and binary NS mergers, will be investigated in a forthcoming study.

%
\textit{Acknowledgments}--- G.G. is grateful to P. Proust for helpful discussions regarding INM properties of N2LO Skyrme EDFs.
G.G. and N.N.S. are also thankful to M. Oertel for insightful comments on the finite-temperature EoS.
This work was supported by the Fonds de la Recherche Scientifique (F.R.S.-FNRS) and the Fonds Wetenschappelijk Onderzoek -- Vlaanderen (FWO) under the EOS Projects nr O022818F and O000422F. 
The present research benefited from computational resources made available on the Tier-1 supercomputers Zenobe and Lucia of the Fédération Wallonie-Bruxelles,
infrastructure funded by the Walloon Region under the grant agreement nr 1117545.
Further computational resources have been provided by the clusters Consortium des Équipements de Calcul Intensif (CÉCI), funded by F.R.S.-FNRS under Grant No. 2.5020.11 and by the Walloon Region. Additionally, resources and services used in this work were provided by the VSC (Flemish Supercomputer Center), funded by the Research Foundation -- Flanders (FWO) and the Flemish Government. We acknowledge EuroHPC Joint Undertaking for awarding us access to MareNostrum5 at BSC, Spain.
W.R., N.C., and S.G. gratefully acknowledge support from the F.R.S.-FNRS.
\appendix

\section{Supplementary material}
\label{app:supl}

The Skyrme energy density functional (EDF) provides the foundation for our nuclear model, expressing the total energy as an integral over local energy densities. These densities are constructed from the nucleon distributions and their spatial variations (i.e., gradients), capturing the complex interplay of the nuclear force.
The EDF is organized by the order of gradients, corresponding to the sensitivity to local momentum and spin-current distributions. The Leading-Order (LO, no gradients) terms describe the bulk properties of nuclear matter. In contrast, the Next-to-Leading-Order (NLO, two gradients) terms refine the description by accounting for surface effects, effective mass, and spin-orbit coupling. The key advancement of our model is the consistent inclusion of Next-to-Next-to-Leading-Order (N2LO, four gradients) terms. This provides a systematic extension of the traditional Skyrme form, offering new degrees of freedom that are essential for simultaneously achieving a realistic equation of state (EoS) for infinite nuclear matter (INM) and high accuracy in the properties of the atomic nuclei. Furthermore, these terms can be viewed as a method to incorporate finite-range physics into the computationally efficient Skyrme framework, while also addressing long-standing spectroscopic issues.

\subsection{The complete N2LO Skyrme EDF}

Using the notation of Ref.~\cite{Ryssens21} for the local densities and currents, the
complete Skyrme EDF is thus a sum of time-even and time-odd terms at each order, as expressed in Eq. (1) of the Letter, with individual components given by

\begin{strip}
\begin{align}
\label{eq:SkTeven:0}
\mathcal{E}^{(0)}_{\rm Sk, e} (\bold{r})
  = & \sum_{t=0,1}
      \Big[ 
      \cc{A}_{t,\textrm{e}}^{(0,1)}    \big( D^{1,1}_t \big)^2
    + \cc{A}^{(0,2)}_{t,\textrm{e}}    \big( D^{1,1}_0 \big)^\alpha 
                                       \big( D^{1,1}_t \big)^2
      \Big] , 
\\[6pt]
\label{eq:SkTeven:2}
\mathcal{E}^{(2)}_{\rm Sk,e} (\bold{r})
  = & \sum_{t=0,1}
      \Big[
      \cc{A}^{(2,1)}_{t,\textrm{e}} D_t^{1,1} \big( \Delta D_t^{1,1} \big)    
    + \cc{A}^{(2,2)}_{t,\textrm{e}} D_t^{1,1} D_t^{(\nabla,\nabla)} 
    + \cc{A}^{(2,3)}_{t,\textrm{e}} \sum_{\mu \nu} C^{1, \nabla \sigma}_{t, \mu \nu}
                                                   C^{1, \nabla \sigma}_{t, \mu \nu} 
     + \cc{A}^{(2,4)}_{t,\textrm{e}} D_t^{1,1} 
           \big(  \nabla \cdot \vec{C}^{1,\nabla \times \sigma}_t \big)
      \Big] , 
\\[6pt]
\label{eq:SkTeven:4}
  \mathcal{E}^{(4)}_{\text{Sk,e}} (\bold{r})
  = & \sum_{t=0,1}
     \Big[
     \cc{A}^{(4,1)}_{t,\textrm{e}} \big(\Delta D^{1,1}_t \big)^2
   + \cc{A}^{(4,2)}_{t,\textrm{e}} D^{1,1}_t \, D^{\Delta, \Delta}_t
   + \cc{A}^{(4,3)}_{t,\textrm{e}} D^{(\nabla, \nabla)}_t D^{(\nabla, \nabla)}_t
   + \cc{A}^{(4,4)}_{t,\textrm{e}} \sum_{\mu\nu} D^{\nabla, \nabla}_{t, \mu\nu}
                                                 D^{\nabla, \nabla}_{t, \mu\nu}      
\nonumber \\[-3pt]
   & \quad
   + \cc{A}^{(4,5)}_{t,\textrm{e}} \sum_{\mu\nu} D^{\nabla, \nabla}_{t,\mu\nu}
                                              \big( \nabla_{\mu} \nabla_{\nu}  D^{1,1}_t \big)
   +  \cc{A}^{(4,6)}_{t,\textrm{e}} \sum_{\mu\nu} C^{1,\nabla\sigma}_{t,\mu \nu} 
      \big( \Delta C^{1,\nabla\sigma}_{t,\mu \nu} \big)
   +  \cc{A}^{(4,7)}_{t,\textrm{e}}  \sum_{\mu\nu\kappa} 
            \big( \nabla_{\mu} C^{1,\nabla\sigma}_{t,\mu \kappa} \big)
            \big( \nabla_{\nu} C^{1,\nabla\sigma}_{t,\nu \kappa} \big)
   +  \cc{A}^{(4,8)}_{t,\textrm{e}} \sum_{\mu\nu} 
       C^{1,\nabla\sigma}_{t,\mu\nu} C^{\Delta,\nabla \sigma}_{t,\mu\nu}
\Big] , 
\\[6pt]
\label{eq:SkTodd:0}
\mathcal{E}^{(0)}_{\text{Sk,o}}(\bold{r})
  = & \sum_{t=0,1}
      \Big[ 
      \cc{A}^{(0,1)}_{t,\textrm{o}}         
                 \vec{D}^{1,\sigma}_t \cdot  \vec{D}^{1,\sigma}_t
    + \cc{A}^{(0,2)}_{t,\textrm{o}} (D^{1,1}_0 )^\alpha \, 
                 \vec{D}^{1,\sigma}_t \cdot  \vec{D}^{1,\sigma}_t
      \Big] , 
\\[6pt]
\label{eq:SkTodd:2}
\mathcal{E}^{(2)}_{\text{Sk,o}}(\bold{r})
  = &  \sum_{t=0,1}
      \Big[
     \cc{A}^{(2,1)}_{t,\textrm{o}}   
           \vec{D}^{1, \sigma}_{t} \cdot \big(  \Delta \vec{D}^{1, \sigma}_{t} \big) 
    + \cc{A}^{(2,2)}_{t,\textrm{o}}  
            \vec{D}^{1,\sigma}_{t} \cdot         \vec{D}^{(\nabla, \nabla)\sigma}_{t} 
    + \cc{A}^{(2,3)}_{t,\textrm{o}}   
           \vec{C}^{1,\nabla}_{t}  \cdot         \vec{C}^{1,\nabla}_{t}  
    + \cc{A}^{(2,4)}_{t,\textrm{o}}    
           \vec{D}_{t}^{1,\sigma}  \cdot  \big( \vnabla \times \vec{C}^{1, \nabla}_t\big)
     \Big] , 
\\[6pt]
\label{eq:SkTodd:4}
\mathcal{E}^{(4)}_{\text{Sk,o}}(\bold{r})
  =  & \sum_{t=0,1}
      \Big[
       \cc{A}^{(4,1)}_{t,\textrm{o}}
           \big(   \Delta \vec{D}^{1,\sigma}_t \big) \cdot
           \big( \Delta\vec{D}^{1, \sigma}_t \big)
  +    \cc{A}^{(4,2)}_{t,\textrm{o}}
                      \vec{D}^{1,\sigma}_t \cdot \vec{D}^{\Delta, \Delta\sigma}_t
  +    \cc{A}^{(4,3)}_{t,\textrm{o}}
             \vec{D}^{(\nabla, \nabla) \sigma}_t  \cdot
             \vec{D}^{(\nabla, \nabla) \sigma}_t
  +    \cc{A}^{(4,4)}_{t,\textrm{o}} \sum_{\mu \nu \kappa}
             D_{\mu \nu \kappa}^{ \nabla, \nabla \sigma}
             D_{\mu \nu \kappa}^{ \nabla, \nabla \sigma}
\nonumber \\[-3pt]
&   \quad
  +    \cc{A}^{(4,5)}_{t,\textrm{o}}   \sum_{\mu \nu \kappa}
             D_{\mu \nu \kappa}^{ \nabla, \nabla \sigma}
         \big( \nabla_{\mu} \nabla_{\nu} D_{\kappa}^{1, \sigma} \big)
  +    \cc{A}^{(4,6)}_{t,\textrm{o}}
                      \vec{C}_t^{1, \nabla} \cdot \big( \Delta \vec{C}_t^{1, \nabla} \big)
  +    \cc{A}^{(4,7)}_{t,\textrm{o}}
                 \big(\vnabla \cdot \vec{C}_t^{1, \nabla}\big)^2
  +    \cc{A}^{(4,8)}_{t,\textrm{o}}
              \vec{C}^{1, \nabla}_t \cdot \vec{C}^{\Delta,\nabla}_t
     \Big] .
\end{align}
\end{strip}

%
%
One of the advantages of this notation is that it directly indicates the gradient
and spin structure of the densities.
The systematic analysis of Ref.~\cite{Ryssens21} provides a non-redundant set of densities with a given number of gradient operators. Since we are interested only in the particle-hole channel of the Skyrme EDF, each derivative order yields two irreducible normal densities, $D$ and $C$. With zero, two, and four gradients, we have ordinary densities,
\begin{align}
D_q^{1,1}, \quad
D^{1,\sigma}_{q,\mu}, \quad
D^{\nabla,\nabla}_{q,\mu\nu} \,, \quad
D^{\nabla,\nabla\sigma}_{q,\mu\nu\kappa}, \quad
D^{\Delta,\Delta}_{q}, \quad
D^{\Delta,\Delta\sigma}_{q,\mu},
\end{align}
while current densities emerge with one and three gradients
\begin{align}
C^{1,\nabla}_{q,\mu}, \quad
C^{1,\nabla\sigma}_{q,\mu\nu}, \quad
C^{\Delta,\nabla}_{q,\mu}, \quad
C^{\Delta,\nabla\sigma}_{q,\mu}. \quad
\end{align}
These are ten objects appearing in Eqs.~\eqref{eq:SkTeven:0}-\eqref{eq:SkTodd:4}\footnote{Note, that $C_{q,\mu}^{1,\nabla\times\sigma}=\sum\limits_{\nu\kappa}^{} \epsilon_{\mu\nu\kappa}C_{q,\nu\kappa}^{1,\nabla\sigma}$.}. To connect with most of the previous studies, limited to the NLO level, we supply the following dictionary~\cite{Ryssens21}:

\begin{equation}
\begin{split}
D_q^{1,1}(\mathbf{r}) &\rightarrow \rho_q(\mathbf{r}) \,, \\
D^{(\nabla,\nabla)}_{q}(\mathbf{r}) &\rightarrow \tau_q(\mathbf{r}) \,,  \\
C^{1,\nabla}_{q,\mu}(\mathbf{r}) &\rightarrow j_{q,\mu}(\mathbf{r}) \,,
\end{split}
\quad\quad
\begin{split}
 D^{1,\sigma}_{q,\mu} (\mathbf{r}) &\rightarrow s_{q,\mu} (\mathbf{r}) \,,\\
 D^{(\nabla,\nabla)\sigma}_{q,\mu\nu\kappa}(\mathbf{r}) &\rightarrow T_{q,\mu}(\mathbf{r}) \,,\\
C^{1,\nabla\sigma}_{q,\mu\nu}(\mathbf{r}) &\rightarrow J_{q,\mu\nu}(\mathbf{r}) \,.\\
\end{split}
\end{equation}

It allows matching with traditionally used time-even quantities: $\rho_q(\mathbf{r})$ being number density, $\tau_q(\mathbf{r})$ - kinetic density, $J_{q,\mu\nu}(\mathbf{r})$ - spin-current density, and their time-odd analogues: $s_{q,\mu} (\mathbf{r})$ - spin density, $T_{q,\mu}(\mathbf{r})$ - kinetic spin density, $j_{q,\mu}(\mathbf{r})$ - current density, respectively. For the link with the higher order terms of Ref.~\cite{Becker17} we refer reader to Ref.~\cite{Ryssens21}. 

The coupling constants determine the strength of each term in the functional. Their values are constrained by the properties of nuclear matter and experimental data on finite nuclei, as detailed in our fitting procedure. 

The N2LO EDFs considered here are constructed from density-dependent effective contact interactions. 
The expressions for the coupling constants of the 
time-even LO and NLO terms of the Skyrme functional, Eqs.~\eqref{eq:SkTeven:0} 
and \eqref{eq:SkTeven:2} are: 
\begin{alignat}{4}
\cc{A}_{t,\rm e}^{(0,1)} =&   
 + \phantom{\tfrac{1}{6}}  C^{+}_{0t}(t_0, x_0) \, , &
\nonumber \\
\cc{A}_{t,\rm e}^{(0,2)} =& 
 + \tfrac{1}{6}    C^{+}_{0t}(t_3, x_3) \, ,  &
\nonumber \\
\cc{A}_{t,\rm e}^{(2,1)} =&  
  -  \tfrac{3}{8} C^{+}_{0t}(t_1, x_1) 
  +  \tfrac{1}{8} C^{-}_{0t}(t_2, x_2) \, , &
\nonumber \\
\cc{A}_{t,\rm e}^{(2,2)} =&
  +   \tfrac{1}{2} C^{+}_{0t}(t_1, x_1) 
  +   \tfrac{1}{2} C^{-}_{0t}(t_2, x_2) \, , & 
\nonumber \\
\cc{A}_{t,\rm e}^{(2,3)} =&  
  -  \tfrac{1}{2} C_{1t}^{+}(t_1,x_1)   
  -  \tfrac{1}{2} C_{1t}^{-}(t_2,x_2)  \, , & 
\nonumber \\
\cc{A}_{0,\rm e}^{(2,4)} =& 
            - \tfrac{1}{2}  W_{\rm 0} -\tfrac{1}{4}  W'_{\rm 0} \, , & \nonumber \\
\cc{A}_{1,\rm e}^{(2,4)} = &  
            - \tfrac{1}{4}  W'_{\rm 0} \, . &
\end{alignat}
The coupling constants for the time-odd terms at LO and NLO are given by
%
\begin{alignat}{4}
\cc{A}_{t,\rm o}^{(0,1)} =& 
 + \phantom{\tfrac{1}{6}}  C^{+}_{1t}(t_0, x_0) \, , &
\nonumber \\
\cc{A}_{t,\rm o}^{(0,2)} =& 
 +  \tfrac{1}{6}   C^{+}_{1t}(t_3, x_3) \, , &
\nonumber \\
\cc{A}_{t,\rm o}^{(2,1)} =&
            - \tfrac{3}{8}  C_{1t}^+ (t_1, x_1)      &
            + \tfrac{1}{8}  C_{1t}^- (t_2, x_2)\, ,  &
\nonumber \\
\cc{A}_{t,\rm o}^{(2,2)} = &  
             + \tfrac{1}{2}  C_{1t}^{+}(t_1,x_1)  & 
             + \tfrac{1}{2}  C_{1t}^{-}(t_2,x_2)\, ,   &
\nonumber \\
\cc{A}_{t,\rm o}^{(2,3)} = & 
            - \tfrac{1}{2}   C^{+}_{0t}(t_1, x_1) &
            - \tfrac{1}{2}   C^{-}_{0t}(t_2, x_2)\, ,   &
\nonumber \\
\cc{A}_{0,\rm o}^{(2,4)} =&
            - \tfrac{1}{2}  W_0  - \tfrac{1}{4}  W'_0 \, , & & 
\nonumber \\
\cc{A}_{1,\rm o}^{(2,4)} =&
            - \tfrac{1}{4}  W'_0\, . & & 
\end{alignat}
The coupling constants appearing in the time-even, N2LO part of the functional 
$\mathcal{E}_{\rm Sk, e}^{(4)}$ of Eq.~\eqref{eq:SkTeven:4}, are
\begin{alignat}{5}
\cc{A}_{t,\rm e}^{(4,1)} = & 
  + \tfrac{3}{16} &  C^{+}_{0t}(t_1^{(4)}, x_1^{(4)})  &
  - \tfrac{1}{16} &  C^{-}_{0t}(t_2^{(4)}, x_2^{(4)}) \, , \nonumber\\
\cc{A}_{t,\rm e}^{(4,2)} =  
 &  + \tfrac{1}{4} & C^{+}_{0t}(t_1^{(4)}, x_1^{(4)})  &
    + \tfrac{1}{4} & C^{-}_{0t}(t_2^{(4)}, x_2^{(4)}) \, , \nonumber\\
\cc{A}_{t,\rm e}^{(4,3)} =  
 & + \tfrac{1}{4} & C^{+}_{0t}(t_1^{(4)}, x_1^{(4)}) &
   + \tfrac{1}{4} & C^{-}_{0t}(t_2^{(4)}, x_2^{(4)}) \, , \nonumber\\
\cc{A}_{t,\rm e}^{(4,4)} =
 & + \tfrac{1}{2} & C^{+}_{0t}(t_1^{(4)}, x_1^{(4)}) &
   + \tfrac{1}{2} & C^{-}_{0t}(t_2^{(4)}, x_2^{(4)}) \, , \nonumber\\
\cc{A}_{t,\rm e}^{(4,5)} =
 & -  \tfrac{1}{2} & C^{+}_{0t}(t_1^{(4)}, x_1^{(4)}) &
   -  \tfrac{1}{2} & C^{-}_{0t}(t_2^{(4)}, x_2^{(4)}) \, , \nonumber\\
\cc{A}_{t,\rm e}^{(4,6)} = 
 & -  \tfrac{1}{4} & C^{+}_{1t}(t_1^{(4)}, x_1^{(4)}) &
   -  \tfrac{1}{4} & C^{-}_{1t}(t_2^{(4)}, x_2^{(4)})\, , \nonumber \\
\cc{A}_{t,\rm e}^{(4,7)} = 
 & -  \tfrac{1}{2} & C^{+}_{1t}(t_1^{(4)}, x_1^{(4)})  &
   -  \tfrac{1}{2} & C^{-}_{1t}(t_2^{(4)}, x_2^{(4)})\, , \nonumber \\ 
\cc{A}_{t,\rm e}^{(4,8)} =  
&  + \phantom{\tfrac{1}{1}} & C^{+}_{1t}(t_1^{(4)}, x_1^{(4)})  &
   + \phantom{\tfrac{1}{1}} & C^{-}_{1t}(t_2^{(4)}, x_2^{(4)})     \,.
\end{alignat}
We have for the coupling constants appearing in the time-odd N2LO energy 
density $\mathcal{E}_{\rm Sk,o}^{(4)}$ of Eq.~\eqref{eq:SkTodd:4}
\begin{alignat}{5}
\cc{A}_{t,\rm o}^{(4,1)} = 
 &          +  \tfrac{3}{16} & C^{+}_{1t}(t_1^{(4)}, x_1^{(4)}) &
            -  \tfrac{1}{16} & C^{-}_{1t}(t_2^{(4)}, x_2^{(4)}) \, , \nonumber \\
\cc{A}_{t,\rm o}^{(4,2)} =  
 &          +  \tfrac{1}{4}  & C^{+}_{1t}(t_1^{(4)}, x_1^{(4)}) &
            +  \tfrac{1}{4}  & C^{-}_{1t}(t_2^{(4)}, x_2^{(4)}) \, , \nonumber\\
\cc{A}_{t,\rm o}^{(4,3)} =  
 &          +  \tfrac{1}{4}  & C^{+}_{1t}(t_1^{(4)}, x_1^{(4)}) &
            +  \tfrac{1}{4}  & C^{-}_{1t}(t_2^{(4)}, x_2^{(4)}) \, , \nonumber \\
\cc{A}_{t,\rm o}^{(4,4)} =
 &          + \tfrac{1}{2}   & C^{+}_{1t}(t_1^{(4)}, x_1^{(4)}) &
            + \tfrac{1}{2}   & C^{-}_{1t}(t_2^{(4)}, x_2^{(4)}) \, ,  \nonumber\\
\cc{A}_{t,\rm o}^{(4,5)} =
 &          - \tfrac{1}{2}   & C^{+}_{1t}(t_1^{(4)}, x_1^{(4)}) &
            - \tfrac{1}{2}   & C^{-}_{1t}(t_2^{(4)}, x_2^{(4)}) \, , \nonumber\\
\cc{A}_{t,\rm o}^{(4,6)} = 
 &          - \tfrac{1}{4}   & C^{+}_{0t}(t_1^{(4)}, x_1^{(4)}) &
            - \tfrac{1}{4}   & C^{-}_{0t}(t_2^{(4)}, x_2^{(4)}) \, , \nonumber\\
\cc{A}_{t,\rm o}^{(4,7)} = 
 &          - \tfrac{1}{2}   & C^{+}_{0t}(t_1^{(4)}, x_1^{(4)})  &
            - \tfrac{1}{2}   & C^{-}_{0t}(t_2^{(4)}, x_2^{(4)}) \, , \nonumber\\ 
\cc{A}_{t,\rm o}^{(4,8)} =  
 &          + \phantom{\tfrac{1}{1}}  & C^{+}_{0t}(t_1^{(4)}, x_1^{(4)})  &
            + \phantom{\tfrac{1}{1}}  & C^{-}_{0t}(t_2^{(4)}, x_2^{(4)})  \,.
\end{alignat}
where the short-hand notation gives the connection with the usual $t,x$ parameters
\begin{alignat}{2}
C^{+}_{00} (t,x) &= +\tfrac{3}{8} t \, , \phantom{+\tfrac{1}{4} tx } & \quad
C^{+}_{01} (t,x) &= -\tfrac{1}{8} t - \tfrac{1}{4} tx           \, , \nonumber \\
C^{+}_{10} (t,x) &= -\tfrac{1}{8} t + \tfrac{1}{4} tx           \, , &
C^{+}_{11} (t,x) &= -\tfrac{1}{8} t \, , \phantom{+ \tfrac{1}{4} tx} \nonumber \\
C^{-}_{00} (t,x) &= +\tfrac{5}{8} t + \tfrac{1}{2} tx           \, , &
C^{-}_{01} (t,x) &= +\tfrac{1}{8} t + \tfrac{1}{4} tx           \, , \nonumber \\
C^{-}_{10} (t,x) &= +\tfrac{1}{8} t + \tfrac{1}{4} tx           \, , &
C^{-}_{11} (t,x) &= +\tfrac{1}{8} t \, .\phantom{+ \tfrac{1}{4} tx } 
\label{eq:CCshort:8}
\end{alignat}
%

\begin{table}[h]
    \centering
\caption{\label{tab:param_skyrme}The BSkG5 parameter set: fifteen parameters determining the self-consistent
mean-field energy $E_{\rm HFB}$, three to the pairing 
functional, and nine determining the correction energy 
$E_{\rm corr}$. For comparison, we include the values of the BSkG4 parameter set~\cite{Grams25}. Note that instead of parameter $x_2$ we list the values of the product $x_2 \, t_2$. }
    \begin{tabular}{c|cc}
        \hline
        \hline
Parameters & \mbox{BSkG4} & \mbox{BSkG5} \\
\hline
    $t_0$         [MeV fm$^3$]               & -2325.45      &  -1873.13\\
$t_1$         [MeV fm$^5$]               & 731.84        &  336.211  \\
$t_2$         [MeV fm$^5$]               & 0.01          &  0.01     \\
$t_3$         [MeV fm$^{3 + 3\alpha}$]   & 14092.79      &12393.856 \\
$t_4$         [MeV fm$^{5 + 3\beta}$]    & -476.32       &  \\
$t_5$         [MeV fm$^{5 + 3\gamma}$]   & 271.19        &  \\
$x_0$                                    & 0.549106      &  0.217333 \\
$x_1$                                    & 2.97317       & -0.0076547 \\
$x_2 t_2$     [MeV fm$^5$]               & -431.435904   & -209.0288873 \\
$x_3$                                    & 0.618431      & 0.06578978 \\
$x_4$                                    & 5.87636       &   \\
$x_5$                                    & 0.353345      &   \\
$t_1^{(4)}$   [MeV fm$^7$]               &               & -12.4982 \\
$t_2^{(4)}$   [MeV fm$^7$]               &               & -34.7836 \\
$x_1^{(4)}$                              &               & -0.09002 \\
$x_2^{(4)}$                              &               & -1.85941 \\
$W_0$         [MeV fm$^5$]               & 122.206       &  125.831 \\
$W_0'$        [MeV fm$^5$]               & 79.840        &  111.514 \\
$\alpha$                                 & 1/5           &  0.3  \\
$\beta$                                  & 1/12          & \\
$\gamma$                                 & 1/4           & \\
\hline
$\kappa_{n}$ [fm$^8$]                    & 123.20        &  148.266  \\
$\kappa_{p}$ [fm$^8$]                    & 129.07        &  147.906   \\
$E_{\rm cut}$ [MeV]                      & 7.919         &  7.799  \\
\hline
$b$                                      & 0.905         &  0.952   \\
$c$                                      & 6.764         &  9.690  \\
$d$                                      & 0.234         &  0.274   \\
$l$                                      & 1.787         &  5.399   \\
$\beta_{\rm vib}$                        & 0.866         &  0.829  \\
\hline
$V_W$         [MeV]                      & -1.411        & -1.868   \\
$\lambda$                                & 560.00        &   315.79  \\
$V_W'$        [MeV]                      & 0.531         &   0.862  \\
$A_0$                                    & 38.174        &   34.927  \\
\hline
\hline
\end{tabular}
\end{table}

We show in Table \ref{tab:param_skyrme} the values of all 27 parameters 
that characterize the BSkG5 model: the first group of 15 parameters determines the Skyrme part of the functional, the second group consists of 3 parameters specifying the pairing terms, the third group of 5 parameters and the final group of four parameters governs the collective correction and the Wigner
energy, respectively. For comparison, Table~\ref{tab:param_skyrme} also 
contains all parameter values of BSkG4~\cite{Grams25}. 
Note that the N2LO-EDF used for BSkG5 does not increase the number of model parameters. BSkG5 has four parameters related to the central terms with four gradients in the Skyrme EDF ($t_1^{(4)}$,$t_2^{(4)}$, $x_1^{(4)}$, $x_2^{(4)}$) which were not present on the NLO form of BSkG4. We, however, removed the density-dependent terms from the extended Skyrme EDF presented in BSkG4, which reduces the number of parameters by six ($t_4$, $t_5$, $x_4$, $x_5$, $\beta$, $\gamma$). Therefore,  BSkG5 has two fewer parameters than BSkG4.

\subsection{Infinite nuclear matter properties}
\label{sec:INM}

In the following, we define the densities used in INM, 
and write the expressions for INM thermodynamic properties with Skyrme N2LO EDF.

Considering a finite cubic volume $V=L^3$ with periodic boundary conditions\footnote{Note that in the thermodynamic limit, the shape of the volume does not matter.}, the single-particle states are plane waves represented by 
\begin{equation}
   \psi_{qs\bold{k}}(\bold{r},\sigma) \equiv  \frac{1}{\sqrt{V}}\exp(i \bold{k \cdot r}) \chi_s^{(q)}(\sigma) \; ,
\end{equation}
where $q=n,p$ for neutrons, protons respectively, 
$s$ the spin projection relative to the quantization axis, and $\chi_s^{(q)}(\sigma)$ denotes Pauli spinor. In the thermodynamic limit, the components $k_\mu=2\pi n_\mu/L$ where $n_\mu=0,\pm 1,\pm 2 \cdots$ of the wave vector $\bold k$ vary quasicontinuously. Discrete summations over $\bold k$ can thus be replaced by integrations as follows
\begin{equation}\label{eq:continuum}
\sum_{\bold{k}} \rightarrow    \frac{V}{(2\pi)^3} \int \,d^3k \; .
\end{equation}

\subsubsection{Local densities in INM}

Introducing the occupation number $n_{qs,\bold{k}}$ of the single-particle states of nucleons species $q$ with momentum $\bold k$ and spin projection 
$s$,
\begin{equation}
n_{qs,\bold{k}} =
\begin{cases}
1 & \text{for } |{\bold{k}}| \leq k_{F,qs} \; ,\\
0 & \text{for } |{\bold{k}}| > k_{F,qs} \; ,
\end{cases}
\end{equation}
which, at zero temperature, is a step function of values 0 and 1 for different values of the momentum $k$ ($k_{F,qs}$ is the Fermi momentum of the nucleon $q$ with spin polarization $s$). 

For practicality, in the present section, we write the densities appearing at LO and NLO levels with the usual notation $ \rho_q =   D^{1,1}_{q} $, and $ \tau_q = D^{(\nabla,\nabla)}_{q}$, respectively. 

The number of particles with spin $s$ and isospin $q$ in an arbitrary volume $V$, 
\begin{eqnarray}
 \rho_{qs}(\bold{r})  &=&  D^{1,1}_{qs} (\bold{r}) = \frac{V}{(2\pi)^3}\int_{\infty}  d^3k  \; n_{qs,\bold{k}} \; \psi_{qs\bold{k}}^{\dag}(\bold{r}) \; \psi_{qs\bold{k}}(\bold{r}) \; , \nonumber \\
 &=& \frac{1}{6\pi^2} k_{F,qs}^3 \; ,
\end{eqnarray}

For unpolarized matter, we can write the density of particles as,
\begin{eqnarray}
 \rho_{q}(\bold{r}) = \sum_{s=\uparrow,\downarrow}\rho_{sq}(\bold{r}) =  \frac{1}{3\pi^2} k_{F,q}^3 \; .
\end{eqnarray}

The kinetic densities $\tau_{qs} (\bold{r}) =  D^{(\nabla,\nabla)}_{qs}(\bold{r})$ appearing at NLO level reads,
\begin{eqnarray}
\tau_{qs} (\bold{r}) &=& \frac{V}{(2\pi)^3} \int_{\infty}  d^3k \; n_{qs\bold{k}} \; [{\nabla} \psi_{qs\bold{k}}(\bold{r}) ]^{\dag} \cdot [{\nabla} \psi_{qs\bold{k}}(\bold{r}) ] \; , \nonumber \\
 &=& \frac{1}{10 \pi^2} \; k_{F,qs}^5 \; .
\end{eqnarray}

We again sum the spins $\uparrow$ and $\downarrow$ and obtain:
\begin{equation}
\tau_{q} (\bold{r})  = \frac{1}{5 \pi^2} \; k_{F,q}^5 . 
\end{equation}
At N2LO, we also need the components of the rank-2 tensor of kinetic density $\tau_{qs,\mu \nu}$\footnote{We drop the $\bold{r}$ dependence of the kinetic density here to lighten the notation.}. 
For a spherical Fermi sphere, the off-diagonal, $\mu \neq \nu $, components of $\tau_{qs,\mu \nu}$, are all zero, as they are an odd function of $k_\mu$ and $k_\nu$.
Since $\sum_\mu \tau_{qs,\mu \mu}  = \tau_{qs} $, the symmetries also imply,
\begin{eqnarray}
\tau_{qs,xx}  =\tau_{qs,yy}  =\tau_{qs,zz} = \frac{1}{3}\tau_{qs} = \frac{1}{30 \pi^2} \; k_{F,qs}^5 \;, 
\end{eqnarray}
such that the diagonal terms of $\tau_{q,\mu \nu}$ for unpolarized matter read
\begin{eqnarray}
\tau_{q,\mu \mu}   =\frac{1}{15 \pi^2} \; k_{F,q}^5. 
\end{eqnarray}
The $D^{\Delta, \Delta}_{q s}$ density appears only in N2LO.
In INM, it corresponds to the higher-order momentum of the distribution function. In coordinate space, it is given by,
\begin{eqnarray}
D^{\Delta, \Delta}_{qs} &=& \frac{V}{(2\pi)^3} \int_{\infty}  d^3k \; n_{qs\bold{k}} \; [{\Delta} \psi_{qs\bold{k}}(\bold{r}) ]^{\dag} \cdot [{\Delta} \psi_{qs\bold{k}}(\bold{r}) ] \; , \nonumber \\
 &=& \frac{1}{14 \pi^2} \; k_{F,qs}^7 \; ,    
\end{eqnarray}
for unpolarized matter, we obtain,
\begin{equation}
D^{\Delta, \Delta}_{q}  = \frac{1}{7 \pi^2} \; k_{F,q}^7 . 
\end{equation}

\subsubsection{Densities at arbitrary asymmetries}
\label{sec:denrho}

Here we write now the densities $\rho_0$, $\tau_0$ and $D^{\Delta, \Delta}_{0}$ in arbitrary asymmetries,
\begin{align}
    \rho_0  = \rho_n + \rho_p \, , \\
    \rho_1 = \rho_n - \rho_p \, ,
\end{align}
where the sum of spin up and down is implicit in each $n/p$ density.
For the kinetic densities, we have,
\begin{align}
    \tau_0 = \frac{3}{5}\beta F_{5/3}^{(0)}\rho_0^{5/3} \, , \\
    \tau_1 = \frac{3}{5}\beta F_{5/3}^{(\tau)}\rho_0^{5/3} \, ,
\end{align}
where $\beta \equiv (\frac{3 \pi^2}{2})^{2/3}$ and \cite{chabanat1997}
\begin{align}
  F^{(0)}_m = \frac{1}{2}\left[ (1 + \eta)^m +(1 - \eta)^m \right] , \\
  F^{(\tau)}_m = \frac{1}{2}\left[ (1 + \eta)^m -(1 - \eta)^m \right] ,
\end{align}
with the isospin asymmetry $\eta = \rho_1 / \rho_0$.

The expressions for the high-order kinetic densities appearing in the N2LO term 
read
\begin{align}
 D^{\Delta, \Delta}_{0} = \frac{3}{7}\beta^2 F_{7/3}^{(0)}\rho_0^{7/3} , \\
 D^{\Delta, \Delta}_1 = \frac{3}{7}\beta^2 F_{7/3}^{(\tau)}\rho_0^{7/3} .
\end{align}

\subsubsection{Thermodynamic properties}
\label{app:thermon2lo}

We define the energy per particle as $e = E/A = \mathcal{E}/\rho_0 $ and write the kinetic densities in terms of $\rho_0$ as explained in Section~\ref{sec:denrho}. 
We consider only unpolarized matter below.

\begin{figure}[htbp]
\begin{center}
\includegraphics[width=1.0\columnwidth]{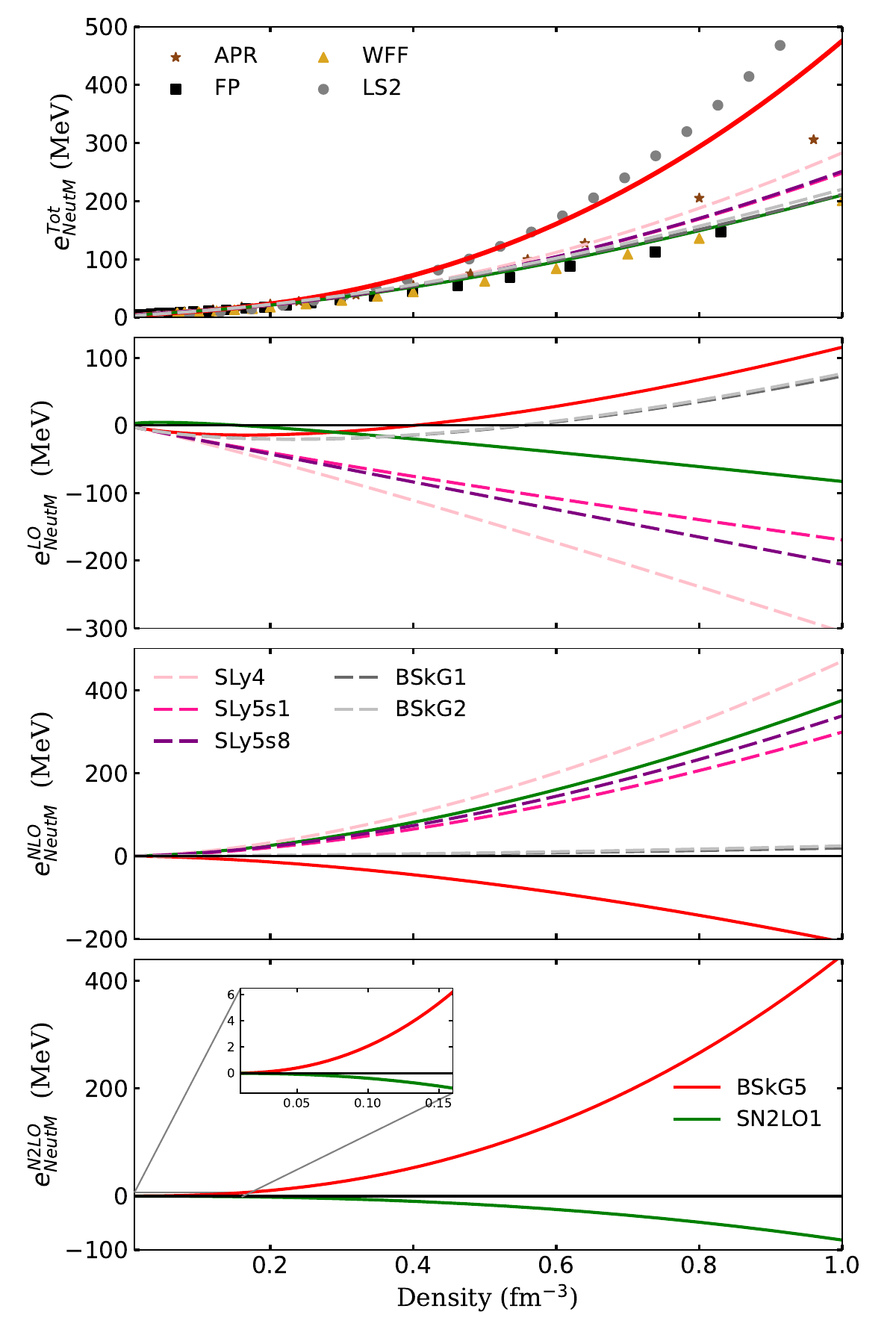}
\end{center}
\caption{
\label{fig:eos} 
Top: Neutron matter energy per particle as a function of the baryon density $\rho_0$, for BSkG5 (solid red), SN2LO1 (solid green), BSkG1 (dashed dark-gray), BSkG2 (dashed light-gray), SLy4 (dashed light-pink), SLy5s1 (dashed dark-pink), and SLy5s8 (dashed purple). 
Markers show the ab-initio calculations of WFF \cite{WFF}, APR \cite{APR}, LS2 \cite{LS2}, and FP \cite{FP}. 
Skyrme N2LO models are shown in solid lines, while standard Skyrme models are shown in dashed lines.
We display the contributions from individual terms for the potential energy, LO (second panel), NLO (third panel), and N2LO (bottom panel). 
The insert plot on the bottom panel shows the $e^{\rm N2LO}_{\rm NeutM}$ at sub-saturation densities.
}
\end{figure}

The energy per particle in INM reads,
\begin{eqnarray}
e^{\rm Tot}(\rho_0,\eta)= e^{\rm kin}+e^{LO} + e^{NLO} +e^{N2LO} \, ,
\label{eq:eTot}
\end{eqnarray}
where the contributions from the individual terms are, 
\begin{eqnarray}
e^{\rm kin} &=& \frac{3\hbar^2}{10 m}F_{5/3}^{(0)}\beta \rho_0^{2/3}    \, ,\\
 e^{LO} &=& [ A_{0,e}^{(0,1)} + A_{0,e}^{(0,2)}\rho_0^{\alpha}  \nonumber \\
 &+&\left( A_{1,e}^{(0,1)} + A_{1,e}^{(0,2)}\rho_0^{\alpha} \right)  \eta^2 ]\rho_0 \, ,  \\
 e^{NLO} &=& \frac{3}{5}\left[ A_{0,e}^{(2,2)}F_{5/3}^{(0)} + A_{1,e}^{(2,2)} \eta F_{5/3}^{(\tau)}\right] \; \beta \rho_0^{5/3} \, ,\\
 e^{N2LO} &=& \frac{3}{7}\left[ A_{0,e}^{(4,2)}F_{7/3}^{(0)} + A_{1,e}^{(4,2)} \eta F_{7/3}^{(\tau)}\right] \; \beta^2 \rho_0^{7/3}  \\
&+& \frac{3}{5}\left[ A_{0,e}^{(4,2)}\left( F_{5/3}^{(0)} \right)^2 + A_{1,e}^{(4,2)} \left( F_{5/3}^{(\tau)}\right)^2 \right] \; \beta^2 \rho_0^{7/3} \, . \nonumber
\label{eq:e2a}
\end{eqnarray}
We display in Fig.~\ref{fig:eos} the energy per particle for pure neutron matter (NeutM), $e^{\rm Tot}_{\rm NeutM}(\rho_0,\eta)=e^{\rm Tot}(\rho_0,\eta=1)$, for the standard NLO (BSkG1~\cite{Scamps21}, BSkG2~\cite{Ryssens22}, SLy4~\cite{Chabanat98}, SLy5s1~\cite{jodon2016}, and SLy5s8~\cite{jodon2016}) and N2LO Skyrme (BSkG5~[this work] and SN2LO1~\cite{Becker17}). 
We show the total energy at the top, and compare our results with the more microscopic calculations of WFF \cite{WFF}, APR \cite{APR}, LS2 \cite{LS2}, and FP \cite{FP}. 
Note that BSkG5 (solid red line) closely follows the stiff behavior of the LS2 results (gray circles), while all other Skyrme parameter sets show a high-density behavior that falls in between what is obtained with APR and FP (or WFF, respectively).

The contributions from the different potential terms are plotted in the second (LO), third (NLO), and bottom (N2LO) panels.
The bottom panel shows that at high densities, the terms entering at the N2LO level are responsible for the stiff EoS of BSkG5, whereas SN2LO1 shows a negative N2LO contribution. Moreover, the insert in the bottom panel shows that the N2LO terms of the SN2LO1 parameterization become negative already at sub-saturation densities.

The pressure is defined as
\begin{equation}
P = \rho_0^{2} \frac{\partial  e^{\rm Tot}}{ \partial \rho_0}\Big|_{\eta}.
\end{equation}

The expression for the pressure then reads,
\begin{eqnarray}
&& P = \frac{\hbar^2}{5 m}F_{5/3}^{(0)}\beta \rho_0^{5/3} \nonumber \\
&+& \left[ A_{0,e}^{(0,1)} + A_{1,e}^{(0,1)} \eta^2 \right]\rho_0^2  \nonumber \\
&+& \left[ A_{0,e}^{(0,2)} + A_{1,e}^{(0,2)} \eta^2 \right](\alpha + 1)\rho_0^{\alpha + 2}  \nonumber \\
&+& \left[ A_{0,e}^{(2,2)}F_{5/3}^{(0)} + A_{1,e}^{(2,2)} \eta F_{5/3}^{(\tau)}\right] \; \beta \rho_0^{8/3} \\
&+& \left[ A_{0,e}^{(4,2)}F_{7/3}^{(0)} + A_{1,e}^{(4,2)} \eta F_{7/3}^{(\tau)}\right] \; \beta^2 \rho_0^{10/3} \nonumber \\
&+& \frac{7}{5}\left[ A_{0,e}^{(4,2)}\left( F_{5/3}^{(0)} \right)^2 + A_{1,e}^{(4,2)} \left( F_{5/3}^{(\tau)}\right)^2 \right] \; \beta^2 \rho_0^{10/3}  \; . \nonumber 
\end{eqnarray}

The incompressibility modulus is defined as,
\begin{eqnarray}
K  &=& \frac{18}{\rho_0} P + 9 \rho_0^2 \frac{\partial^2  e^{\rm Tot}}{\partial \rho_0^2} \, ,
\label{eq:kea}
\end{eqnarray}
we can then write,
\begin{eqnarray}
&& K = 3\frac{\hbar^2}{ m}F_{5/3}^{(0)}\beta \rho_0^{2/3} \nonumber \\
&+& 18 \left[ A_{0,e}^{(0,1)} + A_{1,e}^{(0,1)} \eta^2 \right]\rho_0  \nonumber \\
&+& 9 \left[ A_{0,e}^{(0,2)} + A_{1,e}^{(0,2)} \eta^2 \right](\alpha + 1)(\alpha + 2)\rho_0^{\alpha + 1}  \nonumber \\
&+& 24 \left[ A_{0,e}^{(2,2)}F_{5/3}^{(0)} + A_{1,e}^{(2,2)} \eta F_{5/3}^{(\tau)}\right] \; \beta \rho_0^{5/3} \\
&+& 30 \left[ A_{0,e}^{(4,2)}F_{7/3}^{(0)} + A_{1,e}^{(4,2)} \eta F_{7/3}^{(\tau)}\right] \; \beta^2 \rho_0^{7/3} \nonumber \\
&+& 42 \left[ A_{0,e}^{(4,2)}\left( F_{5/3}^{(0)} \right)^2 + A_{1,e}^{(4,2)} \left( F_{5/3}^{(\tau)}\right)^2 \right] \; \beta^2 \rho_0^{7/3}  \; . \nonumber 
\end{eqnarray}

The skewness is given by the third derivative of the energy per particle and reads,
\begin{eqnarray}
Q &=& 
\frac{12\hbar^{2}}{5m} F^{(0)}_{5/3}\, \beta\, \rho_{0}^{2/3}  \nonumber \\
&& 
+27 
\left[
A^{(0,2)}_{0,e}
+ A^{(0,2)}_{1,e} \eta^{2}
\right]
\alpha (\alpha^{2}-1)\, \rho_{0}^{\alpha+1}   \\
&& 
-6\left[
A^{(2,2)}_{0,e} F^{(0)}_{5/3}
+ A^{(2,2)}_{1,e} \eta F^{(\tau)}_{5/3}
\right]\beta \rho_{0}^{5/3}
\nonumber \\
&&
+ 12\left[
A^{(4,2)}_{0,e} F^{(0)}_{7/3}
+ A^{(4,2)}_{1,e} \eta F^{(\tau)}_{7/3}
\right]\beta^{2}\rho_{0}^{7/3}
\nonumber \\
&&
+\frac{84}{5}\left[
A^{(4,2)}_{0,e}\left(F^{(0)}_{5/3}\right)^{2}
+ A^{(4,2)}_{1,e}\left(F^{(\tau)}_{5/3}\right)^{2}
\right]\beta^{2}\rho_{0}^{7/3} \nonumber.
\end{eqnarray}

We define the symmetry energy as the second derivative of the energy per particle with respect to the asymmetry $\eta$,
\begin{eqnarray}
J = \frac{1}{2} \frac{\partial^2  e^{\rm Tot}}{\partial \eta^2}\Big|_{\rho_0} \; ,
\end{eqnarray}
deriving Eq.~\eqref{eq:eTot}
we obtain,
\begin{eqnarray}
&& J = \frac{\hbar^2}{6 m}F_{-1/3}^{(0)}\beta \rho_0^{2/3}   \nonumber \\
&+&  A_{1,e}^{(0,1)} \rho_0  + A_{1,e}^{(0,2)}  \rho_0^{\alpha + 1} \nonumber \\
&+& \frac{1}{3}\left[ A_{0,e}^{(2,2)}F_{-1/3}^{(0)} + A_{1,e}^{(2,2)} ( \eta F_{-1/3}^{(\tau)} + 3 F_{2/3}^{(0)}) \right] \; \beta \rho_0^{5/3} \\
&+& \frac{2}{3}\left[ A_{0,e}^{(4,2)}F_{1/3}^{(0)} + A_{1,e}^{(4,2)} ( \eta F_{1/3}^{(\tau)} + \frac{3}{2} F_{4/3}^{(0)} ) \right] \; \beta^2 \rho_0^{7/3} \nonumber \\
&+& \frac{2}{3}\left[ A_{0,e}^{(4,2)}\ F_{-1/3}^{(0)} F_{5/3}^{(0)} + A_{1,e}^{(4,2)} F_{-1/3}^{(\tau)}F_{5/3}^{(\tau)} \right] \; \beta^2 \rho_0^{7/3} \nonumber 
\nonumber \\
&+& \frac{5}{3}\left[ A_{0,e}^{(4,2)} \left( F_{2/3}^{(\tau)}\right)^2  + A_{1,e}^{(4,2)} \left( F_{2/3}^{(0)} \right)^2 \right] \; \beta^2 \rho_0^{7/3} .\nonumber 
\label{eq:esym}
\end{eqnarray}

Finally, the slope of the symmetry energy reads,
\begin{eqnarray}
L &=& \frac{\hbar^{2}}{3m} F^{(0)}_{-1/3}\, \beta\, \rho_0^{2/3}  \nonumber \\
&&
+ 3A_{1,e}^{(0,1)}\,\rho_0 + 3
A_{1,e}^{(0,2)}(\alpha+1)\,\rho_0^{\alpha+1} \nonumber \\
&& + \frac{5}{3}
\left[
A_{0,e}^{(2,2)}\,F^{(0)}_{-1/3}
+ A_{1,e}^{(2,2)}\, \eta \, F^{(\tau)}_{-1/3}
\right]
\beta\,\rho_0^{5/3} \\
&&
+ 5\,A_{1,e}^{(2,2)}\,F^{(0)}_{2/3}\,\beta\,\rho_0^{5/3} + 7\,A_{1,e}^{(4,2)}\,F^{(0)}_{4/3}\,\beta^{2}\rho_0^{7/3} \nonumber \\
&&+ \frac{14}{3}
\left[
A_{0,e}^{(4,2)}\,F^{(0)}_{1/3}
+ A_{1,e}^{(4,2)}\,\eta \,F^{(\tau)}_{1/3}
\right]
\beta^{2}\rho_0^{7/3}
\nonumber\\
&&
+ \frac{14}{3}
\left[
A_{0,e}^{(4,2)}\,F^{(0)}_{-1/3}F^{(0)}_{5/3}
+ A_{1,e}^{(4,2)}\,F^{(\tau)}_{-1/3}F^{(\tau)}_{5/3}
\right]
\beta^{2}\rho_0^{7/3}
\nonumber\\
&&
+ \frac{35}{3}
\left[
A_{0,e}^{(4,2)}\left(F^{(\tau)}_{2/3}\right)^{2}
+
A_{1,e}^{(4,2)}\left(F^{(0)}_{2/3}\right)^{2}
\right]
\beta^{2}\rho_0^{7/3} . \nonumber
\end{eqnarray}

\subsubsection{Nucleon effective mass}
\label{sec:effmass}

The effective mass $m^{*}_{q}(k)$ is generally defined from the single-particle energy $\varepsilon_{q}(k)$ as (see, e.g. \cite{JeukenneEtAl1976})\footnote{Note that $\varepsilon_{q}(k)$ is a function of $k$ only here. Therefore  $m^{*}_{q}(k)$ coincides with the so called $k$-effective mass, the $E$-effective mass being equal to the bare mass in this case.} 
\begin{eqnarray}
\label{eq:def_eff_mass}
  \frac{m^{*}_{q}(k)}{m_q}  = \frac{\hbar^2 k}{m_q}\left(\frac{d \varepsilon_{q}}{d k}\right)^{-1}   \; .
\end{eqnarray}
Here $\varepsilon_{q}(k)$ is given by
%
\begin{eqnarray}
  \varepsilon_q =&& F^{1,1}_q ( \rho_q,\rho_{\bar{q}},\tau_q,\tau_{\bar{q}}, D^{\Delta, \Delta}_q, D^{\Delta, \Delta}_{\bar{q}}) \nonumber \\
  +&& F^{\nabla,\nabla}_q ( \rho_q,\rho_{\bar{q}},\tau_q,\tau_{\bar{q}})\; k^2  
 +  F^{\Delta,\Delta}_q (\rho_q,\rho_{\bar{q}})\; k^4,  
\label{eq:spe}
\end{eqnarray}
with,
%
\begin{eqnarray}
F^{1,1}_q &=& 2 [A_{0,e}^{(0,1)} + A_{1,e}^{(0,1)}]\rho_q + 2 [A_{0,e}^{(0,1)} - A_{1,e}^{(0,1)}]\rho_{\Bar{q}} \nonumber \\ 
&+& 2 [A_{0,e}^{(0,2)} + A_{1,e}^{(0,2)}](\rho_q +\rho_{\Bar{q}})^{\alpha}\rho_q  \nonumber \\ 
&+& 2 [A_{0,e}^{(0,2)} - A_{1,e}^{(0,2)}](\rho_q +\rho_{\Bar{q}})^{\alpha}\rho_{\Bar{q}} \nonumber \\
&+& \alpha [A_{0,e}^{(0,2)}(\rho_q +\rho_{\Bar{q}})^2 + A_{1,e}^{(0,2)}(\rho_q -\rho_{\Bar{q}})^2](\rho_q +\rho_{\Bar{q}})^{\alpha - 1} \nonumber \\ 
&+&  [A_{0,e}^{(2,2)} + A_{1,e}^{(2,2)}]\tau_q + [A_{0,e}^{(2,2)} - A_{1,e}^{(2,2)}]\tau_{\Bar{q}} \nonumber \\ 
&+&  [A_{0,e}^{(4,2)} + A_{1,e}^{(4,2)}]D^{\Delta, \Delta}_q + [A_{0,e}^{(4,2)} - A_{1,e}^{(4,2)}]D^{\Delta, \Delta}_{\Bar{q}} \; , \nonumber \\ 
F^{\nabla,\nabla}_q &=& \frac{\hbar^2}{2m} + [A_{0,e}^{(2,2)} + A_{1,e}^{(2,2)}]\rho_q + [A_{0,e}^{(2,2)} - A_{1,e}^{(2,2)}]\rho_{\Bar{q}} \nonumber \\ 
&+& 2 [A_{0,e}^{(4,3)} + A_{1,e}^{(4,3)}]\tau_q + 2[A_{0,e}^{(4,3)} - A_{1,e}^{(4,3)}]\tau_{\Bar{q}} \nonumber \\
&+& 2 [A_{0,e}^{(4,4)} + A_{1,e}^{(4,4)}]\frac{\tau_q}{3} + 2[A_{0,e}^{(4,4)} - A_{1,e}^{(4,4)}]\frac{\tau_{\Bar{q}}}{3} \; , \nonumber \\ 
F^{\Delta,\Delta}_q &=& [A_{0,e}^{(4,2)} + A_{1,e}^{(4,2)}]\rho_q + [A_{0,e}^{(4,2)} - A_{1,e}^{(4,2)}]\rho_{\Bar{q}}  \; .
\end{eqnarray}

The effective mass then reads
\begin{eqnarray}
\frac{m^{*}_{q}(k)}{m_q} = 
\frac{\hbar^2}{2m_q}\left[ F^{\nabla \nabla}_{q} (\rho_q,\rho_{\bar{q}},\tau_q,\tau_{\bar{q}}) + 2\, F_{q}^{\Delta \Delta}(\rho_q,\rho_{\bar{q}}) \, {k}^2  \right]^{-1}\; .
\label{eq:mstar}
\end{eqnarray}
Landau's effective mass is obtained by evaluating $m^{*}_{q}(k)$ at the Fermi momentum $k_{F,q}$. 
Note that the higher-order gradient terms of the N2LO EDF lead to a 
k-dependent effective mass~\eqref{eq:mstar}, a feature naturally found for
finite-range interactions~\cite{Sellahewa12,Ventura92} and in ab-initio theory~\cite{Hassaneen04,Shang20}.


\subsection{Fission properties}

Our description of the fission barriers and spontaneous fission half-lives mentioned in the main part of the paper 
is based on the framework established in Refs.~\cite{ryssens2023,Sanchez2025}; here we summarize only the most relevant aspects. We start by constructing a first potential energy surface (PES) as a function of $\beta_{20}$ and $\beta_{30}$\footnote{The dimensionless multipole moments $\beta_{\ell m}$ characterize the shape of the matter density of the nucleus, see Ref.~\cite{Scamps21} for their definition.}, maintaining for the moment axial symmetry. We then search for the lowest energy path (LEP) that connects the calculated ground state to a configuration at large deformation at equal energy - the classical turning point - with the PyNEB package~\cite{flynn2022}. Based on this path, we construct \emph{a second} PES in the coordinates $\beta_{20}$ and $\beta_{22}$, i.e., we include triaxial deformation, while fixing the octupole deformation $\beta_{30}(\beta_{20})$ to the energetically optimal value on the first PES. We then rely on PyNEB again to find the least action path (LAP) on the new PES, i.e. the path that maximises the (semiclassical) tunneling probability in this two-dimensional collective subspace; aside from the energy, we also provide PyNEB with the full inertia tensor that we consistently calculate from our static configurations within the adiabatic approximation to time-dependent Hartree-Fock-Bogoliubov simulations~\cite{giuliani2018}. We take the local maxima and the local minimum at large deformation as the values for the fission barriers and the excitation energy of the isomer, respectively; empirical RIPL-3 barriers~\cite{Capote09} and spontaneous fission half-lives are generally only known for nuclei in the actinide region that exhibit mostly the traditional double-humped structure such that these definitions are unambiguous. The tunneling probability along the path can straightforwardly be related to the spontaneous fission probability~\cite{giuliani2018,Sanchez2025}. Although crude, our two-step procedure manages to account for three collective degrees of freedom ($\beta_{20}, \beta_{22}, \beta_{30}$) and hence the combined effect of triaxial and octupole deformation on barriers while only incurring part of the cost of building three-dimensional PESs.

\begin{figure}[htbp]
\begin{center}
\includegraphics[width=1.0\columnwidth]{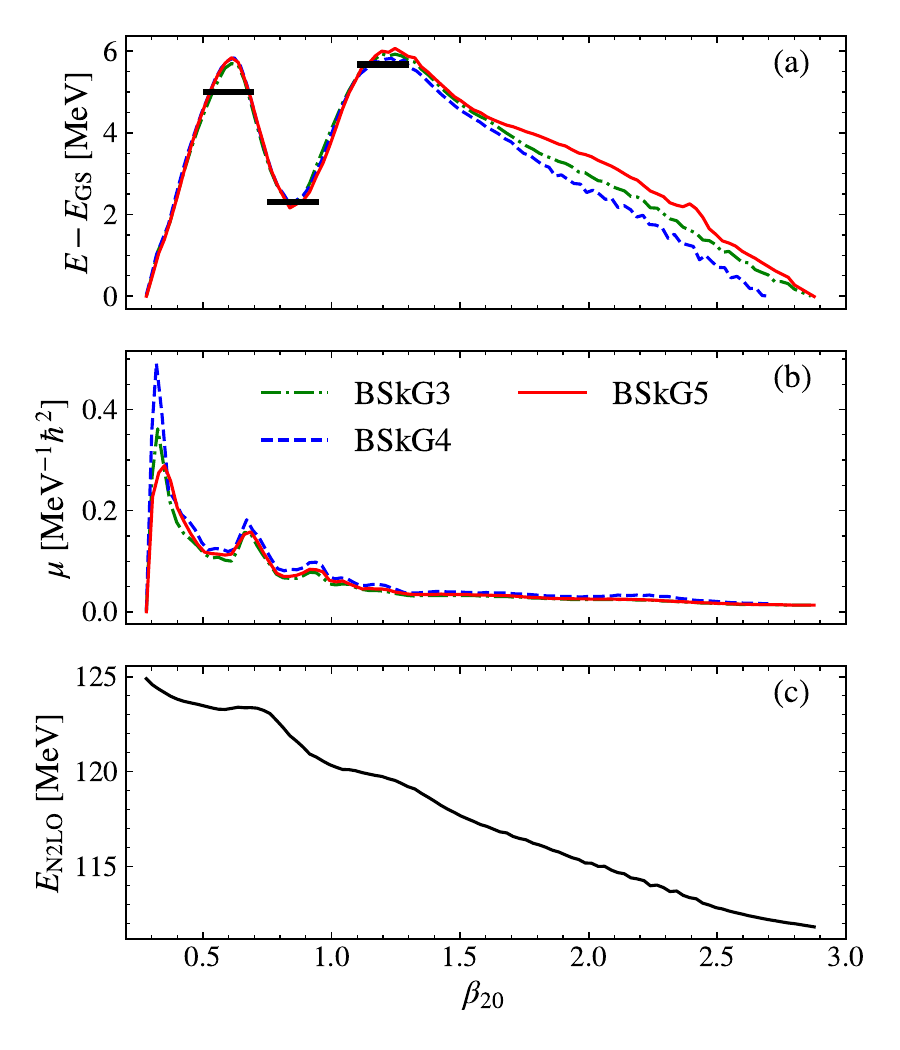}
\end{center}
\caption{
\label{fig:BSkG_U236} (a) One-dimensional fission path and (b) effective inertia as a function of the collective coordinate $\beta_{20}$ for $^{236}$U obtained with BSkG3/4/5.  $E_{\rm GS}$ stands for the ground state energy. Black lines denote the empirical barriers and isomer excitation energy from RIPL-3\cite{Capote09}. (c) Contribution of the N2LO terms of the BSkG5 functional to the total energy along fission path.
}
\end{figure}

As an example, panel (a) of Fig.~\ref{fig:BSkG_U236} shows the total energy (normalized to the ground state) along the LAP as obtained with BSkG3, BSkG4 and BSkG5 as a function of $\beta_{20}$. The overall evolution of the energy as a function of the elongation is similar among all models, although the BSkG5 fission path goes out to slightly higher values of $\beta_{20}$. The empirical barriers of RIPL-3 and the known excitation energy of the isomer are indicated by black horizontal lines. All models perform equally well: they describe the primary barrier height and the excitation energy of the isomer well but overestimate the inner barrier by about one MeV. Panel (b) shows the effective inertia $\mu$ along the LAP: generally speaking,  the shapes of the curves and the locations of the peaks and troughs are similar among all models, another indication that the single-particle structure of all models is comparable. There is a small but noticeable variation in the size of the inertia: BSkG4 calculations produce the largest inertia across nearly the entire fission path. The BSkG3 and BSkG5 curves are quite similar, except near the ground state.  
Because the tunneling probability depends exponentially on $\mu$, this change is sufficient to significantly affect the predicted spontaneous fission lifetime. Panel (c) of Fig.~\ref{fig:BSkG_U236} shows the contribution of all N2LO terms to the total energy along the fission path. As discussed in the main text, the N2LO terms contribute significantly to the total energy, but perhaps more interesting is the fact that this energy contribution does not show significant shell effects and varies rather slowly with deformation. This conclusion is not likely to hold for more general N2LO forms with non-central terms.  Because the total energy remains comparable, the renormalization of the NLO terms in practice counteracts the variation of the N2LO terms.

Our example of $^{236}$U is representative: across the region of actinide nuclei, the results obtained for BSkG5 are comparable to those of BSkG4 in terms of fission paths but with an effective inertia that is generally somewhat smaller, leading to spontaneous fission lifetimes that are on average two orders of magnitude larger. There is a noticeable difference w.r.t. to fission isomers: compared to BSkG4, BSkG5 predictions for isomer excitation energies in odd-mass and odd-odd nuclei are systematically lowered by up to 1.6 MeV, with this maximum reached for $^{240}$Am. 
Although the calculation of the surface tension is beyond the scope of this paper~\cite{proust2022}, the excellent agreement with fission data implies that BSkG5 surface tension is reasonable~\cite{jodon2016}. It also suggests that the correlation between the deformation properties of a Skyrme parameterization and the recipe adopted to account for spurious center-of-mass motion established in Ref.~\cite{dacosta2023} remains valid for N2LO Skyrme EDFs.

\subsection{Rotational moments of inertia}
\label{app:rotMOI}

\begin{figure}
  \centering
 \includegraphics[width=.45\textwidth]{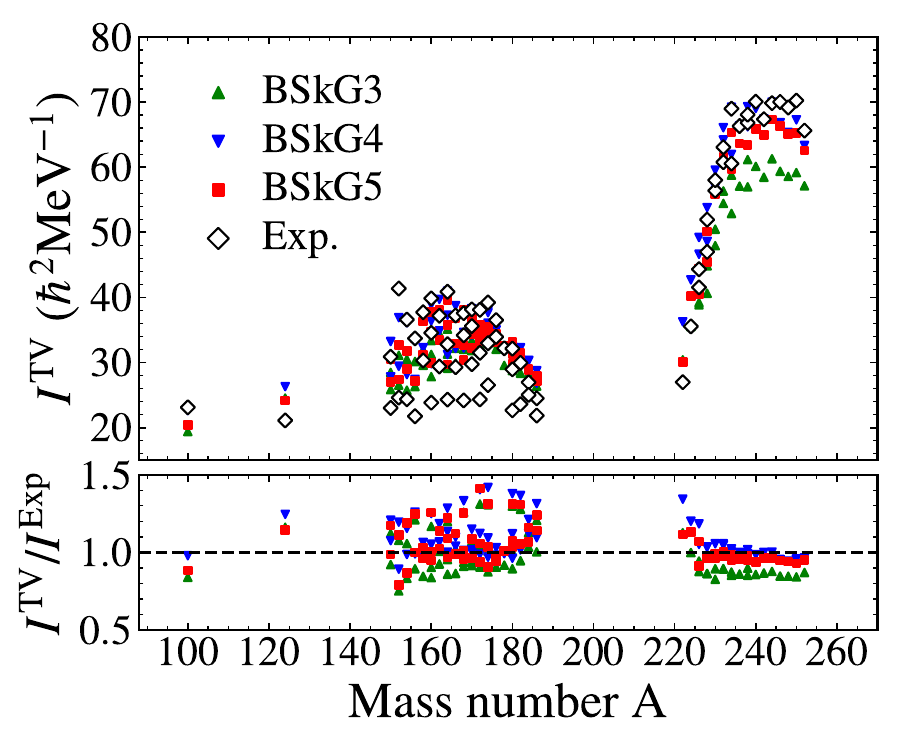}
\caption{Calculated Thouless-Valatin rotational moments of inertia for three BSkG models compared to the experimental data on even-even nuclei of Refs.~\cite{zeng1994,afanasjev2000, raman2001}}.
\label{fig:MOIs}
\end{figure}

We show in Fig.~\ref{fig:MOIs} the kinematical Thouless-Valatin moments of inertia\footnote{Note that these are defined differently than the dynamical moment of inertia discussed in the main paper.} (MoIs)~\cite{thouless1962time} calculated with BSkG3, BSkG4 and BSkG5. To obtain them, we performed cranked HFB calculations with a rotational frequency ${\hbar\omega=0.001}$ MeV starting from the 
ground-state solutions.
We deduced the experimental values of the MoI from known $2^+$ excitation energies \cite{raman2001} belonging to ground-state rotational bands, assuming perfect rigid-rotor behaviour~\cite{afanasjev2000}.
The results for BSkG4 and BSkG5 closely follow the experimental data, particularly for the heaviest nuclei. In contrast, BSkG3 predicts systematically lower values of the MoI, particularly for the heaviest nuclei. 
As discussed in Ref.~\cite{Grams25}, this difference between BSkG3 and BSkG4
is a consequence of the improved interpolation ansatz for the pairing
gaps in INM. As we maintained wholesale our treatment of pairing from BSkG4 to BSkG5, it is encouraging that the overall agreement with MoI data is also conserved, even if BSkG5 produces slightly smaller MoI than BSkG4.

\subsection{Single-particle spectra}
\label{sec:sp-spectra}

\begin{figure}
\centering
\includegraphics[width=.48\textwidth]{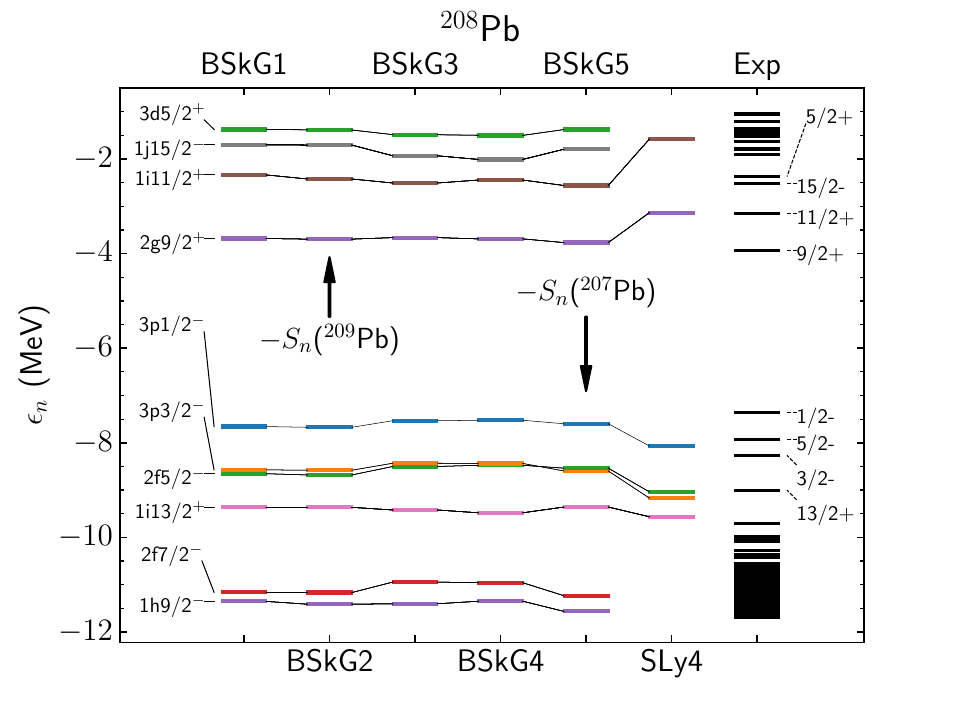}
\caption{Spectrum of the single-particle Hamiltonian for neutrons in $^{208}$Pb as obtained with BSkG1--5 and SLy4~\cite{Chabanat98}.
         Experimental data was extracted from NuDat~\cite{NuDat2025}.}
\label{fig:spec_neutron}
\end{figure}

\begin{figure}
\includegraphics[width=.48\textwidth]{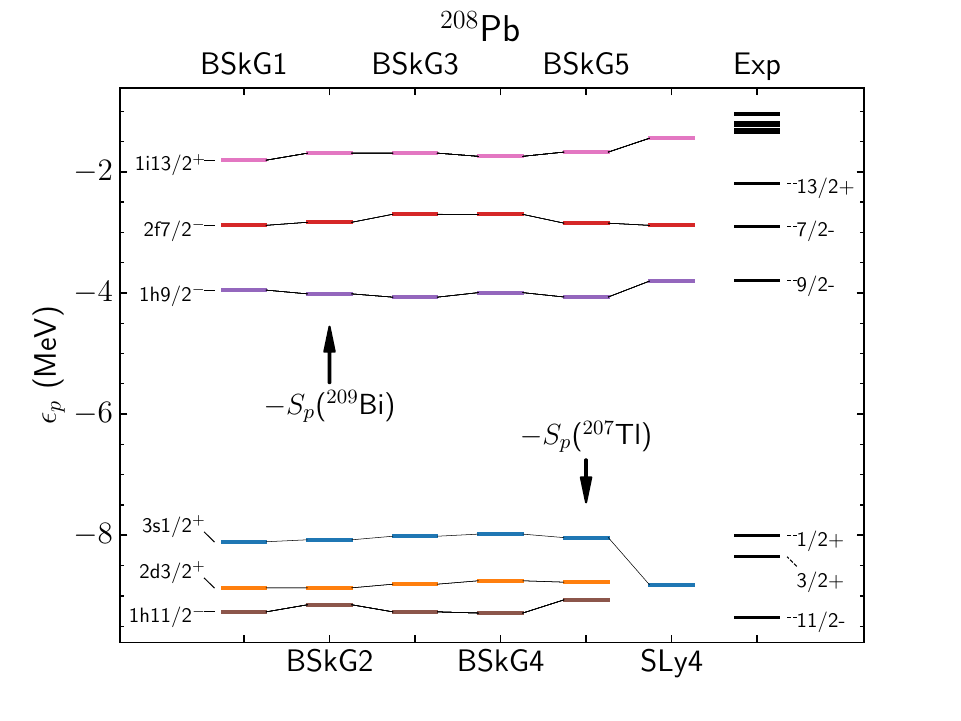}
\caption{Same as Fig.~\ref{fig:spec_neutron}, but for the proton single-particle spectrum of $^{208}$Pb.}
\label{fig:spec_proton}
\end{figure}

As an illustration of the spectroscopic properties of the BSkG models, and BSkG5 in particular, 
in Figs.~\ref{fig:spec_neutron} and~\ref{fig:spec_proton} we present the neutron and proton single-particle spectra of $^{208}$Pb, calculated as eigenvalues of the 
respective single-particle Hamiltonian.
We compare these to experimental data on the one-neutron-separation energies $S_n$
to and from observed low-lying states in $^{207}$Pb and $^{209}$Pb and one-proton-separation energies $S_p$ 
to and from states in $^{207}$Tl and $^{209}$Bi~\cite{NuDat2025}. 
For the lowest observable states, which can be expected to be dominated by a
one-quasiparticle configuration, the separation energy can be reasonably estimated
from the single-particle energies in the even-even neighbor, with a systematic 
error from polarisation and rearrangement effects on the order of a few 100 keV
\cite{rutz1998} that is not relevant for our discussion.

Our most important observation is that the single-particle spectra of $^{208}$Pb obtained are remarkably consistent across all BSkG1--5: with the exception of the nearly degenerate 3p3/2 and 2f5/2 neutron levels, the ordering of levels is identical for all models and even the distances between any two levels remain comparable. This means that all of the models share successes and failures. For example, the ground state spins and parities of $^{207,209}$Pb, $^{207}$Tl and $^{209}$Bi match the quantum numbers of the corresponding levels closest to the Fermi energy;  the energy positioning of the latter is also quite accurate - likely because the binding energies of odd-mass neighbours of $^{208}$Pb were included in the parameter adjustment. On the other hand, the 
neutron intruder orbitals - both the $j15/2$ and the $i13/2$ levels, respectively above and below the Fermi energy - are displaced by roughly one MeV from their experimental location; although we do not go into a deeper analysis here, this is likely to pollute the models predictions for deformed isotopes both heavier and lighter than for which these intruder states become directly relevant to ground state properties.

We also include the spectra of the widely used SLy4 parameterization~\cite{Chabanat98} as a point of reference in Figs.~\ref{fig:spec_neutron} and \ref{fig:spec_proton}: this parameterization too exhibits a comparable ordering of levels, though (i) the separation energies are less well reproduced - likely because none of the neighbours of $^{208}$Pb was included in its fit and (ii) differences between individual levels are too large, likely a consequence of SLy4's comparatively low isoscalar effective mass $m^*/m \sim 0.7$. Further comparisons for other nuclei and published single-particle spectra for NLO Skyrme parameterizations~\cite{Kortelainen14} lead to a similar conclusion: without changing the parameter adjustment protocol, exchanging the NLO form for an N2LO EDF does not lead to a significant shift in single-particle spectra - a similar conclusion was reached for SN2LO1 in Ref.~\cite{Becker17}. This observation does not exclude that N2LO EDFs might improve on the spectroscopic properties of Skyrme NLO EDFs by extending the BSkG5 form with non-central terms, amending the parameter adjustment with suitable observables or a combination of both.

%
%
\subsection{Finite size instabilities and model cutoff}
\label{app:instab}

Tondeur et al.~\cite{tondeur1984} already realized several decades ago that the mean-field equations for Skyrme EDFs do not always admit stable solutions that correspond to finite nuclei. More systematic investigations led to the understanding that EDF parameterisations - whether of Skyrme type or not - can exhibit finite-size instabilities~\cite{lesinski2006,schunck2010,hellemans2012,pastore2015,martini2019} in one or more spin-isospin channels: such models predict that homogeneous INM is unstable with respect to perturbations characterized by a finite wavelength or, equivalently, a transferred momentum $q$ in one or more spin-isospin $(S,T)$ channels. Such finite size instabilities are distinct from infinite-wavelength ($q=0$) instabilities that characterize a phase transition between two homogeneous phases of INM~\cite{ChamelGoriely2010}. Both types of instabilities can be analyzed through linear response theory; the latter reduces to Landau theory for $q=0$.

In practice, a parameterization that is prone to finite size instabilities cannot be used in large-scale models such as ours: when systematically exploring the properties of thousands of nuclei, serious convergence problems are essentially guaranteed to arise~\cite{hellemans2013}. The presence of terms involving higher order derivatives induces higher-order momentum dependencies in the residual interaction matrix elements in linear response theory~\cite{Pastore2015b}, which suggests that the likelihood of encountering finite size instabilities in the fitting process increases when moving from NLO to N2LO EDFs. Avoiding such instabilities is crucial to perform the parameter fit but also to guarantee meaningful results for the final model.

Unfortunately, establishing the stability of a given parameterisation without any doubt is not straightforward. First, an unstable parameterization may lead to divergent calculations only in specific circumstances; examples include the ground state of $^{48}$Ca, whose central isovector density is typically among the largest predicted by EDF models, or cranking calculations at high spin. Numerically, observing (non-)convergence of a given calculation is also no guarantee that a parameterisation is (un)stable; the use of small single-particle bases can mask instabilities~\cite{hellemans2013} while suboptimal solution algorithms for the mean-field equations can diverge for numerical reasons even when the parameterization is stable~\cite{ryssens2019a}.

For NLO Skyrme EDFs, including an empirical criterion based on the linear response of INM in the objective function has been shown to be a practical way to solve this issue~\cite{pastore2015,jodon2016}. The same criterion was employed in an ad-hoc fashion for the construction of SN2LO1~\cite{Becker17}; the limited set of calculations in Ref.~\cite{Ryssens21} indicate that this N2LO parameterisation does not exhibit any finite-size instabilities either. Unfortunately, the linear response code described in Ref.~\cite{becker2015} used for the adjustment of SN2LO1 approximates the impact of the N2LO terms on the effective mass of the nucleons; this is a reasonable approximation for low values of transferred momenta, but deteriorates rapidly with increasing values of $q$, see Fig.~1 in Ref.~\cite{becker2015}. 

Given the absence of a sufficiently flexible linear response tool for N2LO EDFs, we have chosen a more manual approach. In fact, any finite-size instabilities are highly likely to show up as diverging calculations in our fitting protocol for three reasons. First, our coordinate representation guarantees a large single-particle basis. Second, the time-reversal breaking calculations we perform for odd-mass and odd-odd nuclei are sensitive to all $(S,T)$ channels~\cite{Ryssens22}. Third, we  systematically perform calculations for thousands of nuclei such that we probe the entire range of densities relevant to nuclear ground states. We did in fact encounter many unstable parameterizations during the parameter adjustment; when encountered, we manually reoriented the direction of the search in parameter space. We take the existence of the complete set of all masses predicted by BSkG5 as evidence for the stability of the model, and further reinforce our argument by the calculations at high spin for $^{194}$Hg and the linear response calculations for $^{208}$Pb we discuss in the main text.

However, none of these results can guarantee the stability of BSkG5 for arbitrary transferred momenta since all our calculations were performed at a mesh discretisation of $dx \geq 0.8$ fm, which acts as a natural momentum cutoff. As the cost of 3D calculations rises steeply with increasing mesh points,
we employed a spherical HFB code to investigate
the behaviour of BSkG5 at low mesh discretisations. This spherical code
uses the same numerical representation as MOCCa, i.e.\ it
also solves the HFB equations with the two-basis method in a
Lagrange-mesh coordinate-space representation, but in this case
for a radial mesh~\cite{Benderunpublished}. As part of their development, we
have benchmarked both codes in detail for various N2LO parameterizations
and different spherical nuclei and reached agreement to sub-keV level
for the total energy.
Systematic calculations with the spherical code all succeed easily for essentially all value of $dx > 0.3$ fm. Some convergence issues arise for smaller values of $dx$, but it is not excluded that this is due to scaling of the solution algorithm rather than an aspect of BSkG5; the latter seems likely as we were able to converge a 3D calculation with MOCCa, which employs different solution techniques, for $^{48}$Ca even at $dx=0.27$ fm.

Even systematic calculations at low values of $dx$ cannot ensure the stability of the model at very high values of transferred momentum: hence we recommend the use of an explicit ultraviolet cutoff in all future calculations with BSkG5. In coordinate-space representations like ours, the mesh discretisation $dx$ naturally acts as such a cutoff: we cautiously take $dx = 0.4$ fm as the lower bound for calculations with BSkG5, a value that is more than sufficient to converge simulations of the low-energy properties of atomic nuclei~\cite{bulgac2013,Ryssens2015}.
The value of $0.4$ fm is also a reasonable choice from a physical point of view: the degrees of freedom in low-energy nuclear physics are individual nucleons, such that results should not be sensitive to length scales below half the size of the proton.

Our mesh-based cutoff is not directly useful for other types of calculations that do not rely on a discretisation of position-space. We recall the reader that the shortest possible wavelength that can be represented on a uniform mesh of spacing $dx$ in one dimension is given by $\lambda_{\rm min} = 2 dx$. The largest possible wavevector thus obeys $k_{\rm max} = \frac{\pi}{dx}$, implying that a cutoff below $dx=0.4$ fm corresponds to a momentum cutoff above $k_{\rm max} \sim 7.9$ fm$^{-1}$. Not only is this cutoff is far above the relevant momentum scales for nuclear structure, it is also more than sufficient to study neutron stars, as the neutron and proton Fermi momenta reach $k_{\rm max}$ only at $\rho_0 \sim 33$ fm$^{-3}$ (at such unrealistic densities, $\beta$-equilibrated matter is predicted to become symmetric).
\subsection{N2LO EDFs and the microscopic pairing prescription}
\label{app:n2lopair}

Like its predecessors, BSkG3, BSkG4 and several of the older BSk models, BSkG5 relies on the following pairing terms:

\begin{align}
\label{eq:pairing_energy}
E_{\rm pair} &= \frac{1}{4}\sum_{ q=p,n}\int d^3   \bold{r} \, g_{q}(\rho_n, \rho_p) \tilde{\rho}_q^*(\bold{r})\tilde{\rho}_q(\bold{r})\, , \\
g_{q}(\rho_n, \rho_p) &=
V_{q}(\rho_n, \rho_p)  \left[
1 + \kappa_q (\nabla \rho_0)^2 \right]
\, ,
\label{eq:micro_strength}\end{align}
where $\tilde{\rho}_q(\bold{r})$ is the pairing density of species $q=p,n$, the $\kappa_q$ are adjustable parameters~\cite{Grams25} and we employ a cutoff in the form of a Fermi function above the Fermi energy - its cutoff value $E_{\rm cut}$ is an adjustable parameter of the model. We obtain the pairing strengths $V_{q}$ along the lines of Refs.~\cite{chamel2008b,chamel2010}: first, we fix reference pairing gaps in symmetric INM and NeutM those obtained from Brueckner theory~\cite{cao2006} as fitted in Ref.~\cite{goriely2016} and adopt the interpolation scheme of Ref.~\cite{Grams25} to obtain pairing gaps in INM at arbitrary proton-neutron asymmetry $\Delta^{INM}_q(\rho_n, \rho_p)$.
At any given mesh point, we map the density of the simulated nucleus to that of INM at the corresponding asymmetry and invert the pairing problem
to obtain the local pairing strength to match the reference pairing gaps.

The solution strategy of Refs.~\cite{chamel2008b,chamel2010} can still be followed for an N2LO EDF ansatz, but the equations to be solved become slightly more involved.
Adopting the short-hand notation 
$\epsilon_q (k) = U^{(0)}_q (\rho_n, \rho_p) + U^{(2)}(\rho_n, \rho_p) k^2 + U^{(4)}(\rho_n, \rho_p) k^4$ for the single-particle energies 
in (unpolarised) INM - compare with Eq.\eqref{eq:spe} -  the pairing strengths
$V_q(\rho_n, \rho_p)$ 
can be obtained as
\begin{align}
V_{q}(\rho_n, \rho_p) &= - \frac{8 \pi^2}{I_q(\rho_n, \rho_p)} \left(\frac{\hbar^2}{2m^*_q(\rho_n, \rho_p)}\right)^{3/2} \, ,
\label{eq:Vq}
\end{align}
where $I_q$ is the following integral

\begin{strip}
\begin{align}
I_q (\rho_n, \rho_p) &= \int_{0}^{\lambda^{\rm INM}_q + E_{\rm cut}} d \xi \frac{\sqrt{\xi}}{\sqrt{(\xi- \lambda^{\rm INM}_q)^2 + [\Delta^{\rm INM}_{q}(\rho_n, \rho_p)] ^2}}
F\left(\xi,\frac{4U^{(4)}_q (\rho_n, \rho_p)}{[U^{(2)}_q(\rho_n,\rho_p)]^2}\right)
\label{eq:pairing_integral} \, ,
\end{align}
\end{strip}
and the auxiliary function $F(\xi, u)$ is defined as
\begin{align}
F(\xi, u) &= \sqrt{2} \left[ \left( 1 + u \right) \left( 1 + \sqrt{1+u\xi}\right) \right]^{-1/2} \, .
\end{align}
In Eq.~\eqref{eq:Vq}, $m_q^*(\rho_n, \rho_p)$ is the effective mass of species $q$ in INM as defined in Eq.~\eqref{eq:def_eff_mass}.
The integral in Eq.~\eqref{eq:pairing_integral} depends on the pairing cutoff $E_{\rm cut}$
while $\lambda^{\rm INM}_q$ is the reduced Fermi energy in INM, which we estimate using the Fermi wave-number~\cite{chamel2010}:
\begin{align}
\lambda^{\rm INM}_q &\approx U^{(2)}_q k^2_{F,q} + U^{(4)}_q k^4_{F,q}.
\label{eq:}
\end{align}
In the case of an NLO EDF, $U^{(4)}_q = 0$ and these equations reduce to those of Refs.~\cite{chamel2010,Grams23} as they should, though
note the misprints in Ref.~\cite{Grams23} that affect
(a) the exponent of $k_{F,q}$ in the equation defining our approximation for $\lambda^{\rm INM}_q$
and (b) the denominator of the integral which features a symbol $\mu_q$ that should read $\lambda^{\rm INM}_q$.

The presence of the N2LO terms complicates the expression for the integral $I_q$ compared to the NLO case. As before, one can notice that
the integrand is a strongly peaked function near $\xi = \lambda^{\rm INM}_q$ since $\lambda^{\rm INM}_q \gg \Delta^{\rm INM}_q(\rho_n, \rho_p)$.
If the integrand is expanded to first order in $\Delta_q / \lambda_q^{\rm INM}$ and $\Delta_q / E_{\rm cut}$ in the spirit of the weak-coupling approximation,
an analytical result can be obtained
\begin{align}
I_q \approx \lambda_q^{\rm INM} F\left(\lambda_q^{\rm INM},\frac{4U^{(4)}_q (\rho_n, \rho_p)}{[U^{(2)}_q(\rho_n,\rho_p)]^2} \right) \log \left( \frac{4 \lambda^{\rm INM} E_{\rm cut}}{\left[ \Delta_q^{\rm INM} (\rho_n,\rho_p) \right]^2} \right) \, .
\label{eq:analytical_integral}
\end{align}
This approximation is equivalent to assuming a constant density of single-particle states in the gap equations. Making this assumption leads to not-quite-satisfactory results for NLO EDFs~\cite{chamel2010}; we can expect the situation for N2LO EDFs to be worse because of the additional $k^4$-dependence of the single-particle spectrum. In the NLO case, the higher order contributions can be summed analytically such that an accurate approximation of $I_q(\rho_n, \rho_p)$ is available at essentially zero computational cost~\cite{chamel2010}. This is unfortunately not the case for N2LO EDFs: we were unable to obtain an analytical expression valid to higher order because of the presence of $F$ in Eq.~\eqref{eq:pairing_integral}.

Instead, we turned to numerical integration: doing so naively is costly as (i) $I_q(\rho_n, \rho_p)$ has to be evaluated for every iteration at every point on our 3D mesh and (ii) the integrand is a highly-peaked function - especially when $\Delta_{q}^{\rm INM}$ is small, which happens at both low and high densities. To offset the numerical cost, we turned to
so-called tanh-sinh quadrature~\cite{press1992}: the idea is to split the integration interval of Eq.~\eqref{eq:pairing_integral} into
two pieces - $[0, \lambda^{\rm INM}_q$ and $[\lambda^{\rm INM}, \lambda^{\rm INM}_q + E_{\rm cut}]$ - and then apply (i) a linear transformation to recast each into an integral for a variable $x \in [0,1]$ and (ii) a tanh-sinh change of variables $x = \tanh \left( \tfrac{\pi}{2} \sinh (t) \right)$, such that $I_q$ becomes the sum of two indefinite integrals, i.e. $t \in [-\infty, +\infty]$. Because $\frac{dx}{dt}$ is a function that quickly decreases when going away from $t=0$, simple trapezoidal integration in $t$ can be accurate without the need for many collocation points at large $|t|$ and thereby avoiding evaluating the integrand of Eq.~\eqref{eq:pairing_integral} for $\xi \approx \lambda_q^{\rm INM}$. Although detailed error control is possible with this kind of integration algorithm, we opted for a simple strategy: taking equally spaced collocation points $t= -kh, -(k+1)h, \ldots, (k-1)h, kh$ for an integer $k$ and a spacing $h = 0.2$, we grow $k$ until the integrand at the most extreme collocation points diminishes below $10^{-3}$ MeV$^{1/2}$ in absolute size. This threshold and the value of $h$ were tuned manually to compromise between numerical effort and accuracy.

\begin{figure}
  \centering
\includegraphics[width=.4\textwidth]{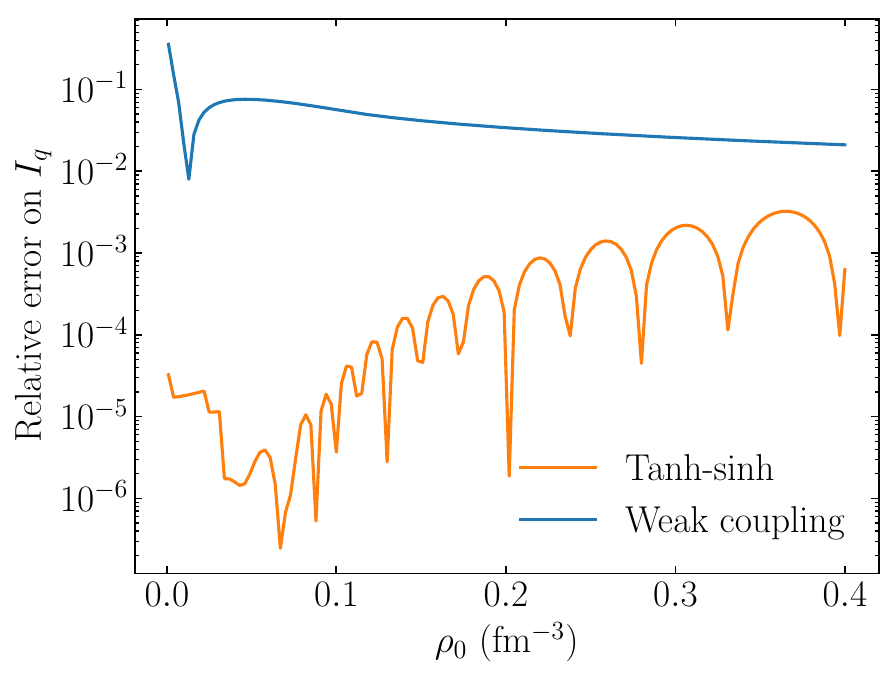}
\caption{Absolute value of the relative error on the calculation of $I_q$ in symmetric INM as a function of the density $\rho_0$. ``Weak coupling'' refers to the analytical approximation Eq.~\eqref{eq:analytical_integral} while ``Tanh-sinh'' refers to the numerical evaluation of Eq.~\eqref{eq:pairing_integral} with our particular implementation of tanh-sinh quadrature; we compare to a brute force solution of Eq.~\eqref{eq:pairing_integral} with relative error less than $10^{-12}$.
}
\label{fig:integral_accuracy}
\end{figure}

In Fig.~\ref{fig:integral_accuracy}, we compare the accuracy of Eq.~\eqref{eq:analytical_integral} compared to our tanh-sinh implementation: even though the analytical approximation is reasonable for higher densities, it can exceed 10\% for densities that are typically encountered during the simulations of finite nuclei. The accuracy of the tanh-sinh integrations is one order of magnitude better at high densities and significantly more accurate at the densities relevant to finite nuclei. Reaching this level of accuracy with our implementation requires less than 60 collocation points at any density in the range $[0,0.4]$ fm$^{-3}$ such that the cost of numerical evaluation of Eq.\eqref{eq:analytical_integral} remains much smaller than many other aspects of our calculations.

\section{Content of the supplementary material files}

Supplementary material includes the complete Skyrme EDF, a table of model parameters, INM expressions, an analysis of the stability of BSkG5, and additional nuclear structure results for the model.

We also provide as supplementary material the data files \newline 
\textsf{Mass\_Table\_BSkG5.dat} and \textsf{Fission\_Table\_BSkG5.dat}.
The former contains the calculated ground-state properties of all nuclei with 
$ 8 \leq Z \leq 118$ lying between the proton and neutron drip lines. The latter
contains the fission barriers and isomer excitation energies calculated for 
all 45 nuclei with $Z \geq 90$ that figure in the RIPL-3 database. 
The contents of both files follow the conventions of the supplementary files of Ref.~\cite{Grams25}.

\bibliographystyle{elsarticle-num} 
\bibliography{ref}
\end{document}